\documentclass[aps,twocolumn,pre,floatfix]{revtex4-2}

\usepackage{booktabs}

\usepackage{xcolor,hyperref}
\hypersetup{
   colorlinks,
   linkcolor={blue!50!black},
   citecolor={blue!50!black},
   urlcolor={blue!80!black}
}

\usepackage{epsfig, amsmath}
\usepackage{bm}
\usepackage{color}
\usepackage{soul,ulem}
\usepackage{hyperref}
\usepackage{footmisc}
\usepackage{wrapfig2,graphicx,enumitem}
\DeclareMathAlphabet{\mathitbf}{OML}{cmm}{b}{it}

\renewcommand{\=}{\!=\!}

\newcommand{\kv}{\mathitbf k}
\newcommand{\uv}{\mathitbf u}
\newcommand{\rv}{\mathitbf r}

\newcommand{\nv}{\mathitbf n}

\newcommand{\calBold}[1]{\mbox{\boldmath${\cal #1}$}}

\begin{document}

\title{Nonlinear phonon dispersion in disordered solids and non-Debye vibrational spectra}
\author{Edan Lerner$^1$}
\email{e.lerner@uva.nl}
\author{Eran Bouchbinder$^2$}
\email{eran.bouchbinder@weizmann.ac.il}
\affiliation{$^1$Institute for Theoretical Physics, University of Amsterdam, Science Park 904, 1098 XH Amsterdam, the Netherlands\\
$^2$Chemical and Biological Physics Department, Weizmann Institute of Science, Rehovot 7610001, Israel}

\begin{abstract}
All solids, whether crystalline or disordered, support elastic wave propagation with a linear dispersion relation in the long-wavelength limit. These waves, corresponding to low-frequency phonons, feature a vibrational density of states that follows Debye's classical model. Deviations from Debye's predictions with increasing frequency can emerge from phonon dispersion nonlinearity and from non-phononic vibrational modes, which exist in non-crystalline solids due to structural disorder. Both nonlinear phonon dispersion in disordered solids and its relative contribution to non-Debye anomalies, most notably manifested by the controversial boson peak, remain poorly understood. Here we show that nonlinear phonon dispersion in a broad range of disordered solids, including elastic networks and various glasses, emerge from a mesoscopic, disorder-induced lengthscale, which also controls wave attenuation. We subsequently use analysis and large-scale computer simulations to quantitatively determine the relative contributions of nonlinear phonon softening and non-phononic vibrations to the onset of non-Debye anomalies and to the boson peak. We show that the relative magnitude of the two contributions strongly depends on the strength of disorder of the solid, e.g., controlled by the thermal history upon glass formation, and that for realistic laboratory glasses both pieces of physics significantly contribute to the boson peak. These findings constitute basic progress in understanding disordered solids.
\end{abstract}

\maketitle

\section{I\lowercase{ntroduction}}

Wave dispersion relations and vibrational spectra are among the most fundamental properties of solids. The former concerns the relation between the frequency $\omega$ of a wave and its wavenumber $k$ (inversely related to its wavelength), quantified through $\omega(k)$. The latter concerns the way in which harmonic vibrational modes, whether wave-like or not, are distributed over frequencies, quantified through the vibrational density of state (VDoS) ${\cal D}(\omega)$. The harmonic vibrational modes of solids generically include long-wavelength phonons, i.e., plane waves, due to global translational symmetry, independently of their microstructure. Since the wavenumber $k$ of such phonons are uniformly distributed, there exists a direct relation between the phonon dispersion relation $\omega(k)$ and the phononic VDoS, involving the group velocity $d\omega(k)/dk$. For ordered, crystalline solids, the phonon dispersion relation and VDoS are well understood, while the corresponding understanding of disordered solids, most notably glasses, is not yet complete.

The origin of this gap is that crystals host only phonons due to their long-range atomic order (periodicity), characterized by a structural lengthscale, the lattice constant (or more generally, the unit cell size)~\cite{kittel2005introduction}. Disordered solids, however, feature additional vibrational excitations on top of phonons~\cite{ramos_book,soft_potential_model_1991,Schober_Laird_numerics_PRB,Gurevich2003,matthieu_PRE_2005,barrat_3d,mw_EM_epl,modes_prl_2016,ikeda_pnas,LB_modes_2019,JCP_Perspective,tanaka_2d_modes_2022} --- which are themselves distorted by disorder~\cite{phonon_widths} --- as well as emergent, disorder-induced lengthscales that are not yet fully understood~\cite{Silbert_prl_2005,pinching_pnas,anomalous_elasticity_soft_matter_2023}. Much of our understanding of crystals follows Debye's celebrated model. Debye focused on the low-frequency (long-wavelength) limit in which acoustic phonons correspond to continuum elastic waves that follow a linear dispersion relation, $\omega(k)\!\simeq\!c\,k$, where $c$ is a relevant wave-speed~\cite{kittel2005introduction}. Assuming the linear dispersion persists to higher wavenumbers/frequencies, he derived the well-known Debye's VDoS of phonons, ${\cal D}_{_{\rm D}}\!(\omega)\=A_{_{\rm D}}\omega^2$ in three dimensions. While Debye's $\omega^2$ law cannot be valid at high frequencies, Debye ensured consistency with the total number of phonons by defining an upper frequency cutoff to his VDoS, known as Debye's frequency $\omega_{_{\rm D}}$, implying $A_{_{\rm D}}\=3/\omega_{_{\rm D}}^3$.

The breakdown of Debye's $\omega^2$ law is related to softening nonlinearity in the phonon dispersion relation $\omega(k)$, which emerges when the wavenumber $k$ becomes comparable to the crystal lattice constant, upon which phonons soften with increasing $k$ (decreasing wavelength). This softening implies that ${\cal D}(\omega)/\omega^2$, known as the reduced VDoS, goes above Debye's plateau level $A_{_{\rm D}}$. In fact, the non-Debye VDoS will feature cusp singularities associated with points at which $d\omega(k)/dk\=0$, known as Van Hove singularities~\cite{van_hove_1953}. The reduced VDoS has also been extensively studied for numerous glasses using experimental vibrational spectra~\cite{ramos_book}. It is universally observed that ${\cal D}(\omega)/\omega^2$ goes above the corresponding Debye's plateau level $A_{_{\rm D}}$ at small frequencies $\omega$ and generically features a peak in the THz range, known as the boson peak (BP)~\cite{sokolov_1986,ramos1997quantitative,wischnewski1998neutron,surovtsev2000inelastic,perez1997low,surovtsev2003density,parisi_boson_peak_2003,grigera2003phonon,wyart_epl_2005,monaco2006density,baldi2023vibrational,tanaka_boson_peak_2008,Schirmacher_2013_boson_peak,eric_boson_peak_emt,gonzalez2022understanding,nature_physics_boson_peak_2025,baldi_prx_2026}.
\begin{figure*}[ht!]
  \includegraphics[width = 0.75\textwidth]{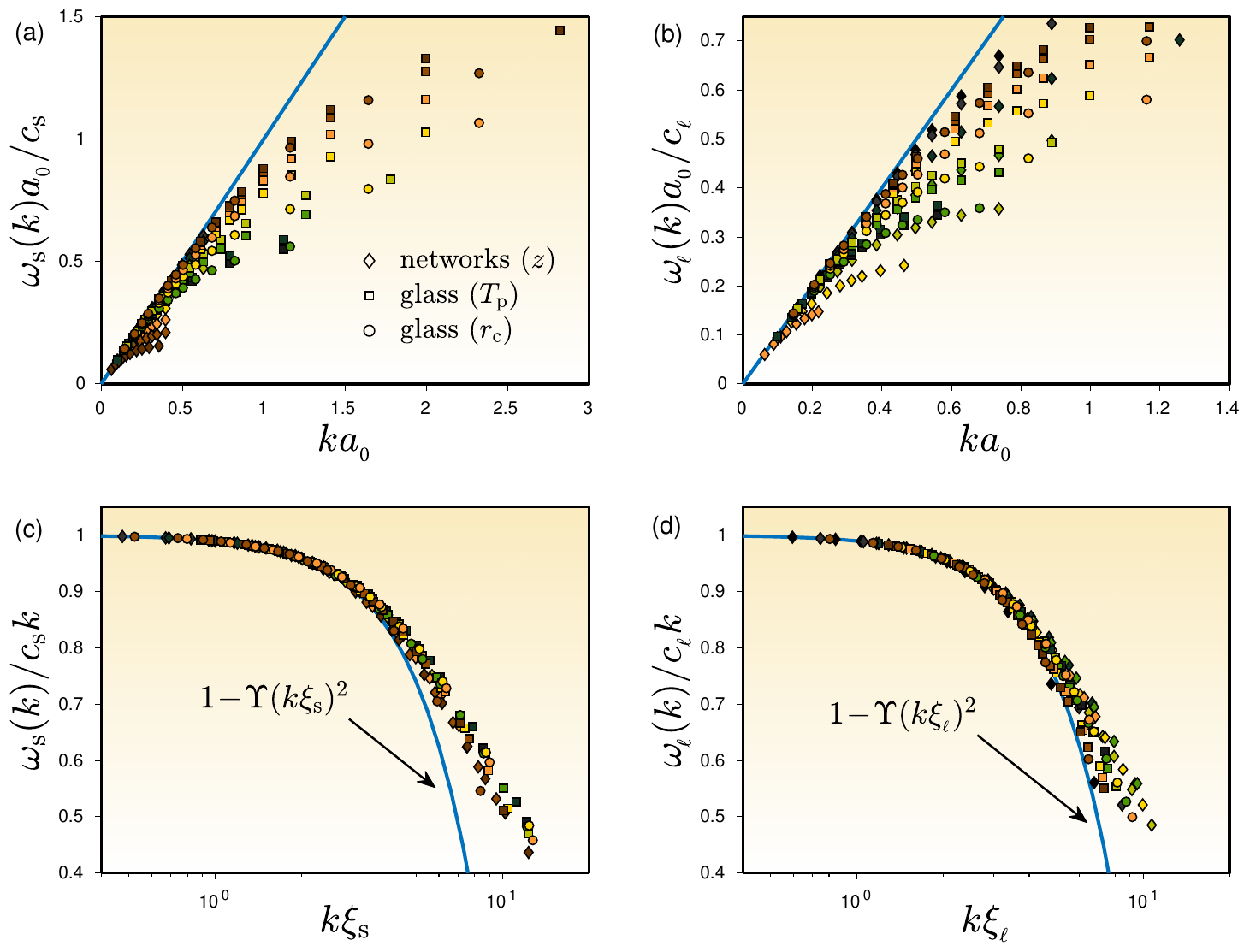}
  \caption{\footnotesize {\bf Phonon dispersion relations of disordered solids}. (a) The shear (transverse) phonon dispersion relation $\omega_{\rm s}(k)$ obtained using the `imposed-wave method' (see Appendix~\ref{sec:appendix_imposed_wave} for details) for 21 computer disordered solids of three different classes (9 disordered elastic networks of different topologies quantified by an average degree of connectivity $z$, 8 polydisperse glasses of different thermal histories upon glass formation quantified by $T_{\rm p}$ and 4 binary glasses of different strengths of interatomic attractive forces quantified by $r_{\rm c}$, see text and Appendix~\ref{sec:appendix_computer_models}). $\omega_{\rm s}$ is normalized by $c_{\rm s}/a_{_0}$, where $c_{\rm s}$ is the shear wave-speed, and the wavenumber $k$ is normalized by $1/a_{_0}$. The linear approximation $\omega_{\rm s}(k)\!\simeq\!c_{\rm s}k$ is marked by the blue line. (b) The same as panel (a) but presenting $\omega_{_{\ell}\!}(k)$ for longitudinal phonons (sound), where $c_{_{\ell}}$ is the longitudinal wave-speed. (c) The same data presented in panel (a) but here plotted as $\omega_{\rm s}(k)/c_{\rm s}k$ vs.~$k\xi_{\rm s}$ on a semi-log scale, where the disorder-induced length $\xi_{\rm s}$ is obtained for each disordered solid by a fit to $1\!-\!\Upsilon(k\xi_{\rm s})^2$ with $\Upsilon\!=\!1/96$ (see blue line and text for additional details). (d) The same as panel (c) but using the data of panel (b) for longitudinal phonons. Here, the disorder-induced length is $\xi_{_{\ell}}$.}
  \label{fig:dispersion}
\end{figure*}

It is now well established that glasses host low-frequency, non-phononic quasi-localized modes~\cite{JCP_Perspective} (in addition to phonons), which contribute to the observed non-Debye vibrational spectra and to the BP~\cite{Gurevich2003,boson_peak_2d_jcp_2023}. Yet, the relative contributions of nonlinear phonon dispersion and non-phononic vibrations to non-Debye spectra are not known. A central obstacle to addressing this fundamental question is that nonlinear phonon dispersion in glasses, as well as the associated disorder-induced lengthscales, are poorly understood due to the presence of structural disorder. Moreover, the effects of the latter on wave attenuation, yet another basic phenomenon in disordered solids, are not well understood. Therefore, a major challenge is to first understand nonlinear phonon dispersion in non-crystalline solids and its similarities/differences compared to crystals, then the relations to wave attenuation, and finally to determine the relative contributions of nonlinear phonon softening and non-phononic vibrations to deviations from Debye's $\omega^2$ law across a broad range of disordered solids.

\vspace{-0.5cm}
\section{N\lowercase{onlinear phonon dispersion in disordered solids}}
\begin{figure*}[ht!]
  \includegraphics[width = 0.75\textwidth]{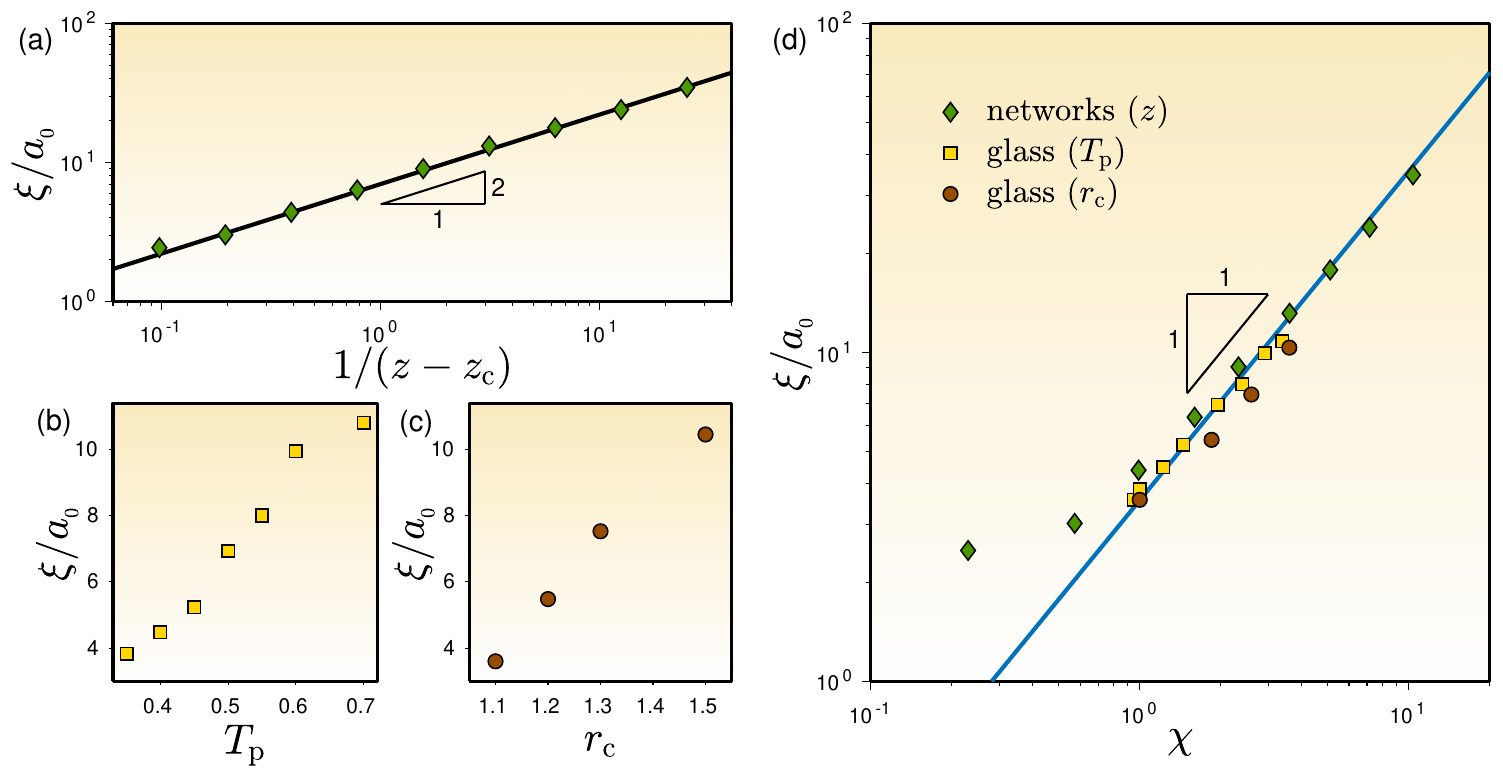}
  \caption{\footnotesize {\bf A disorder-induced dispersion lengthscale}. (a) $\xi/a_{_0}$ (as obtained in Fig.~\ref{fig:dispersion}) vs.~$1/(z\!-\!z_{\rm c})$ for the disordered elastic networks considered therein. The superposed line and the power-law triangle highlight the scaling relation $\xi(z)\!\sim\!1/\sqrt{z\!-\!z_{\rm c}}$, over the entire range of $z$ values used, covering more than two orders of magnitude in $\delta{z}\!=\!z\!-\!z_{\rm c}$. (b) $\xi/a_{_0}$ (as obtained in Fig.~\ref{fig:dispersion}) vs.~$T_{\rm p}$ for the polydisperse glasses considered therein, formed under a wide range of thermal histories. (c) $\xi/a_{_0}$ (as obtained in Fig.~\ref{fig:dispersion}) vs.~$r_{\rm c}$ for the binary glasses considered therein, featuring different strengths of interatomic attractive forces. (d) $\xi/a_{_0}$ vs.~$\chi$ for all disordered solids considered in panels (a)-(c), using the same symbols and colors. $\chi$ is a dimensionless disorder measure that quantifies mesoscopic fluctuations of the shear modulus $\mu$, see text for additional details. The superposed line and the power-law triangle highlight the excellent agreement with the prediction in Eq.~\eqref{eq:xi_chi}, see text for additional discussion.}
  \label{fig:length}
\end{figure*}
The vibrational modes of a low-temperature solid are the eigenfunctions of the Hessian $\bm{\mathcal{H}}$, i.e., the matrix of second derivatives of the potential energy with respect to the particle coordinates, and the corresponding eigenvalues are $\omega^2$. In ordered, crystalline solids phonons of wavenumber $k$ are exact eigenfunctions of $\bm{\mathcal{H}}$. The crux of the challenge to compute the phonon dispersion relation $\omega(k)$ in disordered, non-crystalline solids is that this is not the case due to the presence of structural disorder and of non-phononic vibrations, implying that a unique identification of the frequency $\omega$ of a phonon of wavenumber $k$ is not generally simple.

To overcome this difficulty, we developed a technique termed the `imposed-wave method' (see Appendix~\ref{sec:appendix_imposed_wave}), which allows to assign to each phonon of admissible wavevector $\kv$ in a disordered solid a unique frequency $\omega$. The method is quantitatively validated against established benchmarks and allows to compute the nonlinear phonon dispersion $\omega(k\=|\kv|)$ of computer models of disordered solids. In Fig.~\ref{fig:dispersion}a-b, we present $\omega(k)$ for a large number of disordered solids of three different classes, each characterized by a parameter that is varied over a broad range (see Appendix~\ref{sec:appendix_computer_models} and below).

The first class corresponds to disordered elastic networks, composed of relaxed Hookean springs, featuring both positional and topological (i.e., degree of connectivity) disorder. We control the average coordination number $z$, from very close to the critical Maxwell threshold $z_{\rm c}\=6$ (in three dimensions) to well above it. Second, we employ repulsive polydisperse glasses amenable to very deep supercooling, which is quantified by the temperature $T_{\rm p}$ at which the system goes out of equilibrium. The variation of $T_{\rm p}$ represents a wide range of glass thermal histories, including realistic laboratory conditions. We also employ repulsive binary glasses. Finally, we employ binary glasses characterized by a family of interatomic potentials, featuring both repulsion and attraction, in which the strength of the attractive part is continuously adjustable through a parameter $r_{\rm c}$. These isotropic disordered solids feature shear (transverse) $\omega_{\rm s}(k)$ and longitudinal (sound) $\omega_{\ell}(k)$ dispersion relations.

In Fig.~\ref{fig:dispersion}a, we present $\omega_{\rm s}(k)$ for 21 disordered solids and the corresponding $\omega_{\ell}(k)$ are presented in Fig.~\ref{fig:dispersion}b, using the average interatomic distance $a_{_0}\!\equiv\!(V/N)^{1/3}$ ($N$ is the number of nodes/particles and $V$ is the volume) to nondimensionalize $k$. The phonon dispersion relations in Fig.~\ref{fig:dispersion}a-b showcase the capabilities and utility of the `imposed-wave method' and demonstrate that $\omega(k)$ generically features softening nonlinearity, i.e., it bends over below the linear approximation $\omega(k)\!\simeq\!ck$.

\vspace{-0.5cm}
\section{A \lowercase{disorder-induced dispersion lengthscale}}

The observed nonlinear softening of $\omega(k)$ is qualitatively similar to the corresponding crystalline softening, which is semi-quantitatively described by the standard dispersion relation~\cite{kittel2005introduction}
\begin{equation}
    \omega(k) = \frac{4\,c}{a_{_0}}\sin\!\left(\frac{k\,a_{_0}}{4}\right) \ ,
\label{eq:dispersion_1D_lattice}
\end{equation}
where $a_{_0}$ is not just the average interatomic distance, but actually the exact lattice constant. Here, a deviation from $\omega(k)\!\simeq\!ck$ emerges when $k$ becomes comparable to the microscopic/atomistic inverse structural length $1/a_{_0}$, which immediately raises the fundamental question of whether the nonlinear phonon softening observed in Fig.~\ref{fig:dispersion}a-b for disordered solids is also due to $a_{_0}$, therein the average interatomic distance, or due to another lengthscale that emerges from the presence of disorder.

To address this basic and important question, we take inspiration from Eq.~\eqref{eq:dispersion_1D_lattice} and assume that a general phonon dispersion relation $\omega(k)$ is characterized by a single lengthscale $\xi$, such that it can be expressed as
\begin{equation}
    \omega(k)=f(k\xi)\,ck \ ,
\label{eq:nonlinear_dispersion}
\end{equation}
where $f(x)$ is an even dimensionless function of the dimensionless argument $x\!\equiv\!k\xi$, featuring $f(x\!\to\!0)\!\to\!1$. The question then is whether the dispersion-related length $\xi$ identifies with $a_{_0}$ or not.

To put things on equal footing with crystals, we expand Eq.~\eqref{eq:dispersion_1D_lattice} up to the leading-order (cubic) nonlinearity and present it in the form $\omega(k)/ck\=1-\Upsilon x^2 + {\cal O}(x^4)$, where $\Upsilon\=1/96$ and $\xi\=a_{_0}$. In Fig.~\ref{fig:dispersion}c-d, we use this $\Upsilon$ value to extract $\xi$ from the leading-order (cubic) nonlinearity using the results of Fig.~\ref{fig:dispersion}a-b. We find $\xi\!\gg\!a_{_0}$ for all disordered solids considered, marking the emergence of a mesoscopic lengthscale that does not exist in crystals. We show that $\xi$ is $N$ independent (see Fig.~\ref{fig:fake_phonons}c) and that $\xi_{\rm s}\!\sim\!\xi_{\ell}$ (see Fig.~\ref{fig:wave_speed_validation}), hence hereafter we report only the former (with the notation $\xi$).

Next, we report the dependence of $\xi$ on the control parameters of the three classes of the employed disordered solids. In Fig.~\ref{fig:length}a, we plot $\xi/a_{_0}$ vs.~$1/(z\!-\!z_{\rm c})$ (as obtained in Fig.~\ref{fig:dispersion}) for the disordered networks, demonstrating a remarkable scaling relation $\xi(z)\!\sim\!1/\sqrt{z\!-\!z_{\rm c}}$, likely related to~\cite{Silbert_prl_2005,atsushi_core_size_pre,anomalous_elasticity_soft_matter_2023}. In Fig.~\ref{fig:length}b, we plot $\xi/a_{_0}$ vs.~$T_{\rm p}$ and in Fig.~\ref{fig:length}c, $\xi/a_{_0}$ vs.~$r_{\rm c}$ (both using $\xi$ as obtained in Fig.~\ref{fig:dispersion}). The results in Figs.~\ref{fig:length}a-c clearly reveal the disorder-induced nature of $\xi$ as it is well established that increasing $1/(z\!-\!z_{\rm c})$ in networks, and $T_{\rm p}$ or $r_{\rm c}$ in glasses, systematically increases the degree of disorder. Note that $\xi/a_{_0}$ reaches ${\cal O}(10)$ in glasses, while significantly larger values are observed in disordered networks (cf.~Fig.~\ref{fig:length}a), reflecting the fluidization (unjamming) phase transition occurring as $z\!\to\!z_{\rm c}$~\cite{liu2011jamming}.

\vspace{-0.35cm}
\section{R\lowercase{elations to mesoscopic elasticity fluctuations and wave attenuation}}

The results presented in Fig.~\ref{fig:length}a-c indicate that the dispersion-associated length $\xi$ monotonically increases with the solid's degree of disorder, varied through very different physical parameters and procedures. This qualitative similarity raises the question of whether these diverse physical procedures to vary structural disorder and their effect of $\xi$ can be quantitatively unified. To explore this possibility, one should consider generic and dimensionless quantifiers of disorder in solids.

We focus on a disorder measure, denoted as $\chi$, which quantifies the ratio between the fluctuations of the shear modulus $\mu$ and its mean over mesoscopic lengthscales~\cite{jcp_letter_scattering_2021,phonon_widths2,karina_chi_paper_2023}. It is closely related to the `disorder parameter' of the Heterogeneous Elasticity Theory (HET) of Schirmacher and coworkers~\cite{schirmacher2006thermal,Schirmacher_prl_2007,Schirmacher_2013_boson_peak}, to be further discussed below. Before considering the possible relations between $\chi$ and $\xi$ as the level of disorder is varied by different control parameters, we first broaden the discussion to include yet another basic wave-related phenomenon in disordered solids.

Waves in disorder-free solids propagate indefinitely. Yet, in the presence of disorder, low $k$ waves/phonons undergo attenuation at a rate $\Gamma(k)$ as they propagate, a phenomenon known as Rayleigh scattering with a $\Gamma(k)\!\sim\!k^4$ scaling for small $k$. Recently, the so-called Rayleigh-Klemens relation~\cite{klemens1951thermal} has been rederived, taking the form $\Gamma(k)\!\simeq\!\chi^2 (c k/\omega_{_{\rm D}})^2 \,{\cal D}_{_{\rm D}}\!(c k)\,\omega_{_{\rm D}}^2$~\cite{scattering_jcp,phonon_widths2}, where ${\cal D}_{_{\rm D}}\!(\omega)\=3\omega^2\!/\omega_{_{\rm D}}^3$ and $\omega_{_{\rm D}}$ are Debye's VDoS and frequency, respectively, introduced above.

The above $\Gamma(k)$ appears to be a leading-order expansion of a more general relation that bears similarity to the nonlinear wave dispersion of Eq.~\eqref{eq:nonlinear_dispersion}, i.e., of the form
\begin{equation}
    \Gamma(k) \simeq g(k\xi)\,{\cal D}_{_{\rm D}}\!(c k)\,\omega_{_{\rm D}}^2 \ .
\label{eq:attenuation_rate}
\end{equation}
Here, ${\cal D}_{_{\rm D}}\!(c k)\,\omega_{_{\rm D}}^2$ is a disorder-free rate and $g(x)$ is an even dimensionless function of $x\=k\xi$ that quantifies the effect of disorder, where we hypothesize that the very same disorder-induced dispersion length $\xi$ appears therein. The crucial difference between Eq.~\eqref{eq:nonlinear_dispersion} and Eq.~\eqref{eq:attenuation_rate} is that $g(x\!\to\!0)\!\to\!0$ in the latter, while $f(x\!\to\!0)\!\to\!1$ in the former. That is, as discussed above, while waves in disorder-free continua (where $\xi\=0$) feature a linear dispersion, they experience no attenuation.

Consequently, the leading-order expansion of $g(x)$ follows $g(x)\!\sim\!x^2$, which indeed gives rise to Rayleigh scattering for $x\=k\xi\!\ll\!1$ (recall that ${\cal D}_{_{\rm D}}\!(c k)\!\sim\!k^2$). Comparing then the leading-order expansion of Eq.~\eqref{eq:attenuation_rate} to the above-stated Rayleigh-Klemens relation, we obtain
\begin{equation}
    \xi/a_{_0} \sim \chi \ ,
\label{eq:xi_chi}
\end{equation}
where $\omega_{_{\rm D}}\!\sim\!c/a_{_0}$ has been used. While the relation between $\xi$ and $\chi$ in Eq.~\eqref{eq:xi_chi} has been obtained in the context of wave attenuation, we will test it using the dispersion results presented above, as explained next.

The results in Fig.~\ref{fig:length}a-c present the variation of $\xi$ in the three classes of disordered solids with $1/(z\!-\!z_{\rm c})$, $T_{\rm p}$ and $r_{\rm c}$, as discussed extensively above. To test Eq.~\eqref{eq:xi_chi}, we computed for each of these disordered solids the dimensionless disorder quantifier $\chi$~\cite{phonon_widths2}, and plotted $\xi/a_{_0}$ vs.~$\chi$ in Fig.~\ref{fig:length}d. The results reveal excellent agreement with the prediction in Eq.~\eqref{eq:xi_chi}, except for disordered network with large $z$, where $\xi$ becomes comparable to the microscopic length $a_{_{0}}$.

Our discussion so far, both in relation to nonlinear wave dispersion and to wave attenuation, highlighted the existence of a mesoscopic, disorder-induced lengthscale $\xi$, its variation with diverse control parameters in different classes of disordered solids and how it modifies to leading order the continuum physics of waves in disorder-free solids. It is precisely in this regime where the aforementioned HET~\cite{schirmacher2006thermal,Schirmacher_prl_2007,Schirmacher_2013_boson_peak} is expected to be valid and indeed its leading-order predictions are fully consistent --- in the scaling sense --- with our findings.

Specifically, HET introduces a dimensionless `disorder parameter', denoted as $\gamma$, which has been previously shown to follow $\gamma\!\sim\!\chi^2$~\cite{phonon_widths2}. It then predicts that both the leading-order wave dispersion nonlinearity and wave attenuation rate scale with the square of a length proportional to $\sqrt{\gamma}\,a_0$, in agreement with Eq.~\eqref{eq:xi_chi} and its related findings. Next, we consider disordered solids phenomena that continuum approaches, supplemented with mesoscopic disorder, fall short of accounting for.

\vspace{-0.35cm}
\section{T\lowercase{he onset of non-}D\lowercase{ebye vibrational spectra:} P\lowercase{hononic and non-phononic contributions}}

\begin{figure*}[ht!]
  \includegraphics[width = 1\textwidth]{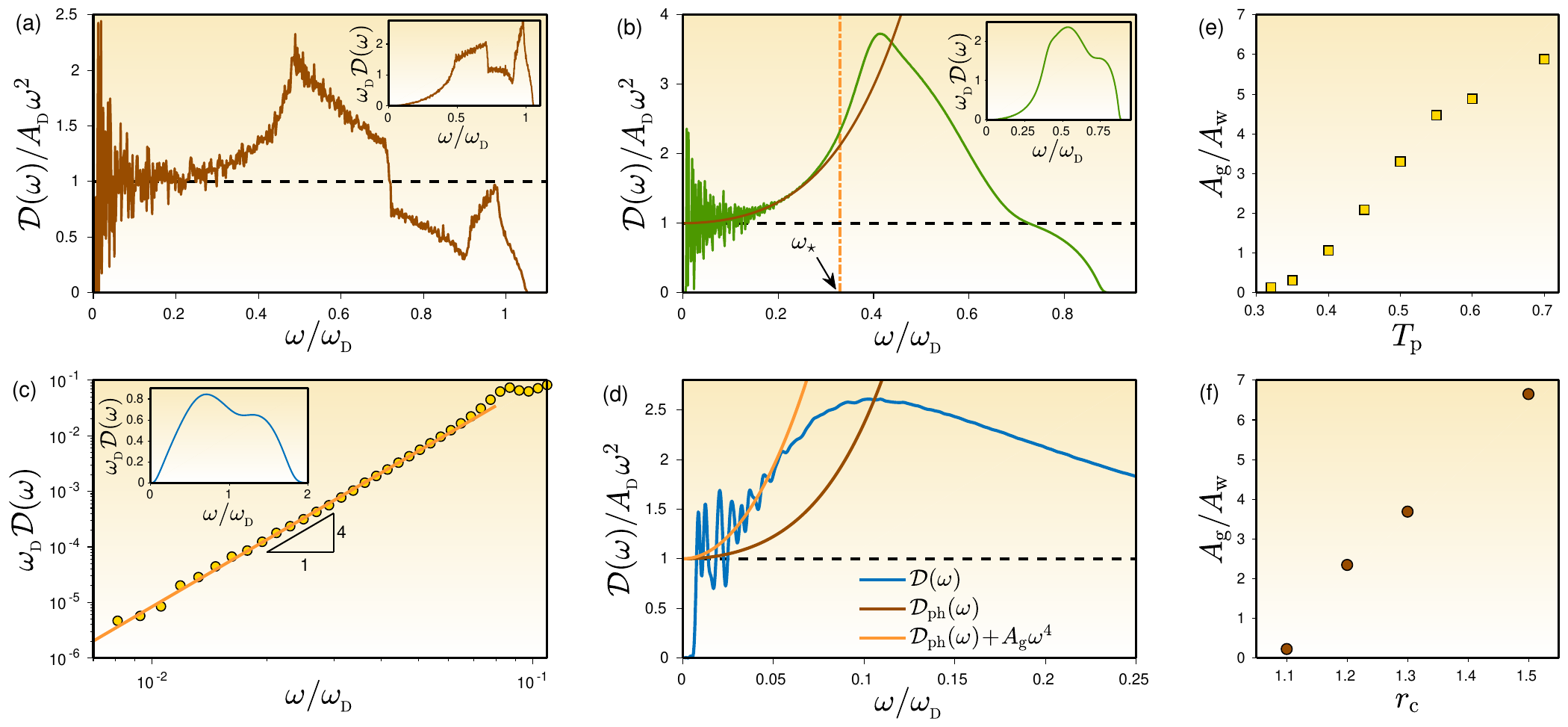}
  \caption{\footnotesize {\bf The onset of non-Debye vibrational spectra}. (a) The reduced VDoS ${\cal D}(\omega)/A_{_{\rm D}}\omega^2$ vs.~$\omega/\omega_{_{\rm D}}$ for a monoatomic face-centered cubic (FCC) crystal. Debye's plateau level, corresponding to unity in this presentation, is marked by the horizontal dashed line. The reduced VDoS features a non-Debye behavior with increasing frequency $\omega$ and a peak corresponding to a Van Hove singularity/cusp, see text for discussion. (inset) The VDoS ${\cal D}(\omega)$ used in the main panel. (b) The same as panel (a), but for a disordered elastic network. The frequency $\omega_\star$ above which non-phononic vibrations emerge (see text and Appendix~\ref{sec:appendix_omega_star}) is marked by the vertical dashed-dotted line. The reduced VDoS features a non-Debye behavior and a boson peak with increasing frequency $\omega$, see text for discussion. The onset of a non-Debye behavior is quantitatively predicted by the phononic VDoS of Eq.~\eqref{eq:phonon_VDoS}, corresponding to the superposed brown curve. (c) The VDoS $\omega_{_{\rm D}}{\cal D}(\omega)$ vs.~$\omega/\omega_{_{\rm D}}$ for a small hyperquenched glass composed of $N\!=\!4$K particles, see text and Appendix~\ref{sec:appendix_BIPL} for details. The superposed solid line and the power-law triangle highlight the agreement with the universal gapless $\sim\!\omega^4$ tail of the non-phononic VDoS of Eq.~\eqref{eq:QLMs_VDoS_tail}, which also allows to robustly extract the amplitude $A_{\rm g}$, see text for discussion. (inset) The VDoS ${\cal D}(\omega)$ for the very same hyperquenched glass, but 1000 times larger, composed of $N\!=\!4$M particles. (d) The same as panel (b), but for the glass whose VDoS is shown in the inset of panel (c). The phononic VDoS of Eq.~\eqref{eq:phonon_VDoS}, corresponding to the superposed brown curve, falls short of accounting for the onset of the non-Debye behavior. Yet, when the non-phononic VDoS of Eq.~\eqref{eq:QLMs_VDoS_tail}, with $A_{\rm g}$ extracted in panel (c), is added --- see superposed orange curve and legend --- excellent agreement with the onset of the non-Debye behavior is observed. (e). The ratio $A_{\rm g}/A_{\rm w}$ vs.~$T_{\rm p}$ for the glasses analyzed in Fig.~\ref{fig:length}b (and also in Fig.~\ref{fig:dispersion}), where $A_{\rm g}$ is extracted as in panel (c) and $A_{\rm w}$ is the amplitude of the $\omega^4$ contribution to ${\cal D}_{\rm ph}(\omega)$ in Eq.~\eqref{eq:phonon_VDoS}. (f) The same as panel (e), but vs.~$T_{\rm p}$ for the glasses analyzed in Fig.~\ref{fig:length}c (and also in Fig.~\ref{fig:dispersion}). See text for discussion.}
  \label{fig:onset}
\end{figure*}
We so far discussed how disorder affects the dispersion of low-frequency phonons and their attenuation with increasing wavenumber, and the associated disorder-induced lengthscale. Since low-frequency vibrational modes play important roles in a broad range of physical phenomena, a basic property of interest is how their number varies with frequency $\omega$, quantified by the VDoS ${\cal D}(\omega)$. As discussed above, the phononic contribution to ${\cal D}(\omega)$ for $\omega\!\to\!0$ is given by Debye's VDoS ${\cal D}_{_{\rm D}}\!(\omega)\=3\omega^2\!/\omega_{_{\rm D}}^3$, which is based on a linear dispersion.

All known solids, crystalline and non-crystalline, feature more vibrational modes with increasing $\omega$ than predicted by Debye, commonly quantified by the reduced VDoS $\omega_{_{\rm D}}^3{\cal D}(\omega)/3\omega^2$~\cite{ramos_book}. It approaches unity for $\omega\!\to\!0$ but increases with $\omega$ over some range of frequencies. If a solid features only phononic vibrational modes, then the emergence of a non-Debye VDoS is exclusively accounted for by a softening nonlinearity in $\omega(k)$. This is the case since the number of phonons in any branch (i.e., shear/transverse and sound/longitudinal) reads~\cite{kittel2005introduction}
\begin{equation}
    n_{\rm ph}(\omega) = \frac{V k^2}{2\pi^2} \left| \frac{dk}{d\omega} \right| \ ,
\label{eq:number_phonons}
\end{equation}
where $k\=k(\omega)$ is the inverse of the dispersion relation $\omega(k)$. For a linear dispersion, $\omega(k)\!\simeq\!c k$, Debye's $\sim\!\omega^2$ VDoS is recovered. Any nonlinearity in $\omega(k)$, however, gives rise to a non-Debye VDoS.

To set a reference for what follows, we computed the VDoS ${\cal D}(\omega)$ of a monoatomic face-centered cubic (FCC) crystal and present its reduced VDoS ${\cal D}(\omega)/A_{_{\rm D}}\omega^2$ in Fig.~\ref{fig:onset}a, which indeed increases above unity with increasing $\omega$, exclusively due to a nonlinear phonon dispersion as in Eq.~\eqref{eq:dispersion_1D_lattice}. In fact, the reduced VDoS reveals a peak in the form of a cusp, corresponding to a vanishing derivative of $\omega(k)$, known as a Van Hove singularity~\cite{van_hove_1953}.

The results in Fig.~\ref{fig:onset}a are conceptually clear, i.e., a non-Debye VDoS exclusively emerges from a nonlinear phonon dispersion $\omega(k)$. If it were also the case for disordered solids, then one would have used the nonlinear phonon dispersion of Eq.~\eqref{eq:nonlinear_dispersion}, with its associated disorder-induced lengthscale $\xi$, to compute the VDoS following Eq.~\eqref{eq:number_phonons}. Specifically, the boson peak (BP) in the reduced VDoS would have been then entirely explained as a counterpart of a Van Hove singularity, exclusively emerging from $\omega(k)$~\cite{Monaco_prl_2011,chumakov2016relation}.

It is now well established, however, that disordered solids feature non-phononic vibrations in addition to phonons~\cite{JCP_Perspective}, implying that non-Debye anomalies cannot be exclusively accounted for by $\omega(k)$. This raises a fundamental question: what are the relative contributions of $\omega(k)$ and non-phononic vibrations to non-Debye spectra in disordered solids? Since we computed $\omega(k)$ for a broad range of disordered solids above and as there are well-developed techniques to extract non-phononic vibrations in these systems, we are in a position to quantitatively address this basic question.

To this aim, we consider Eq.~\eqref{eq:nonlinear_dispersion} up to the leading-order nonlinearity and use Eq.~\eqref{eq:number_phonons} to compute the leading-order phononic corrections to Debye's VDoS, obtaining (see Appendix~\ref{sec:appendix_A6})
\begin{equation}
    {\cal D}_{\rm ph}(\omega) \simeq
     A_{_{\rm D}}\,\omega^2 + A_{\rm w}\,\omega^4 + A_6\,\omega^6 + {\cal O}(\omega^8) \ .
\label{eq:phonon_VDoS}
\end{equation}
Here, the first term is Debye's VDoS (with $ A_{_{\rm D}}\=3\,\omega_{_{\rm D}}^{-3}$) and the other terms correspond to non-Debye contributions due to the leading-order ($k^3$) nonlinearity in $\omega(k)$. We have $A_{\rm w}\=5\,\omega_{\rm w}^{-5}$, involving a characteristic frequency $\omega_{\rm w}\!\sim\!c/(a_{_{0}}^{3/5}\,\xi^{2/5})$ (Supplementary Text) that depends on $\xi$, to be contrasted with Debye's frequency, $\omega_{_{\rm D}}\!\sim\!c/a_{_0}$, which is independent of it. Finally, the sub-dominant $A_6\,\omega^6$ term is included (see Appendix~\ref{sec:appendix_A6}).

We first apply Eq.~\eqref{eq:phonon_VDoS} to a disordered elastic network, known to feature non-phononic vibrations with a gapped distribution, i.e., they appear only above a minimal $z$-dependent frequency $\omega_\star$~\cite{mw_EM_epl}. Therefore, if $\omega_\star$ is sufficiently large, then Eq.~\eqref{eq:phonon_VDoS} is predicted to fully account for the onset of a non-Debye VDoS in this case. This prediction is verified in Fig.~\ref{fig:onset}b, involving no free parameters, hence strongly supporting our procedure for extracting $\omega(k)$ in disordered solids.

We next apply Eq.~\eqref{eq:phonon_VDoS} to glasses, which suffer frustration during glass formation (unlike disordered elastic networks that lack frustration) and hence feature a gapless distribution of non-phononic vibrations~\cite{JCP_Perspective}. These are known to be quasi-localized in space, in qualitative contrast to spatially-extended phonons, and to follow a VDoS ${\cal D}_{\!_{\rm G}\!}(\omega)$ with a universal power-law tail~\cite{modes_prl_2016,modes_prl_2018,modes_prl_2020,ikeda_pnas,LB_modes_2019,JCP_Perspective}
\begin{equation}
    {\cal D}_{\!_{\rm G}\!}(\omega) \simeq A_{\rm g}\,\omega^4 \ ,
\label{eq:QLMs_VDoS_tail}
\end{equation}
for small $\omega$, with no lower cutoff in the thermodynamic limit. The glassy (non-phononic) $A_{\rm g}\,\omega^4$ VDoS clearly competes with the $A_{\rm w}\,\omega^4$ term in Eq.~\eqref{eq:phonon_VDoS} at the onset of non-Debye spectra, where the ratio $A_{\rm g}/A_{\rm w}$ controls their relative importance.

The amplitude $A_{\rm g}$ can be cleanly extracted using well-developed techniques, one of them involves small system sizes such phonons are pushed to higher frequencies~\cite{modes_prl_2016,modes_prl_2018,modes_prl_2020,ikeda_pnas,LB_modes_2019,JCP_Perspective}, exposing the gapless VDoS of Eq.~\eqref{eq:QLMs_VDoS_tail}. This technique is applied in Fig.~\ref{fig:onset}c for a hyperquenched glass, i.e., a glass formed by a very rapid quench, such that $A_{\rm g}$ is obtained. The VDoS ${\cal D}(\omega)$ for the very same glass, but 1000 times larger, is presented in the inset and the corresponding reduced VDoS is shown in Fig.~\ref{fig:onset}d. Superposing then ${\cal D}_{\rm ph}(\omega)$ of Eq.~\eqref{eq:phonon_VDoS}, with no free parameters, shows that the contribution from $\omega(k)$ falls short of accounting for the onset of deviations from Debye's prediction in this case. Adding then the non-phononic VDoS of Eq.~\eqref{eq:QLMs_VDoS_tail}, with $A_{\rm g}$ as extracted in Fig.~\ref{fig:onset}c, leads to excellent quantitative agreement with the non-Debye onset.

The results of Fig.~\ref{fig:onset}d demonstrate that for the considered glass the onset of a non-Debye spectrum is dominated by non-phononic vibrations, where in this case $A_{\rm g}/A_{\rm w}\=3.8$. We repeated this analysis for a broad range of glasses, as a function of both $T_{\rm p}$ and $r_{\rm c}$, and report the results in Fig.~\ref{fig:onset}e-f. It is observed that $A_{\rm g}/A_{\rm w}$ is quite significantly larger than unity at large disorder states, i.e., large $T_{\rm p}$ and $r_{\rm c}$, similarly to Fig.~\ref{fig:onset}d. Yet, at small disorder states, i.e., in stable glasses, the situation is reversed and $A_{\rm g}/A_{\rm w}$ becomes significantly smaller than unity.

These results may indicate that in many laboratory glasses the onset of non-Debye VDoS may be dominated by $\omega(k)$, not by non-phononic vibrations. Specifically, the present authors analyzed experimental vibrational spectra in~\cite{experimental_2024} and observed an $\omega^4$ non-Debye contribution, which was attributed to non-phononic vibrations. The above results appear to indicate that this interpretation is not valid, i.e., that the $\omega^4$ non-Debye contribution may likely emerge from $\omega(k)$. This interpretation may be consistent with another recent example~\cite{baldi_prx_2026}.

The above results about the onset of non-Debye spectra leave the central question of the relative contributions of the two pieces of physics under consideration to the BP, at yet higher frequencies, widely open. The important point to note is that it is known that while the gapless power-law tail of ${\cal D}_{\!_{\rm G}\!}(\omega)$ in Eq.~\eqref{eq:QLMs_VDoS_tail} is very sensitive to the glassy state of disorder, ${\cal D}_{\!_{\rm G}\!}(\omega)$ features a peak beyond the power-law tail, at higher $\omega$'s, which is less sensitive to the glass thermal history~\cite{cge2_jcp2020,moriel2024boson}. Consequently, we next turn our attention to the BP.
\begin{figure*}[ht!]
  \includegraphics[width = 0.75\textwidth]{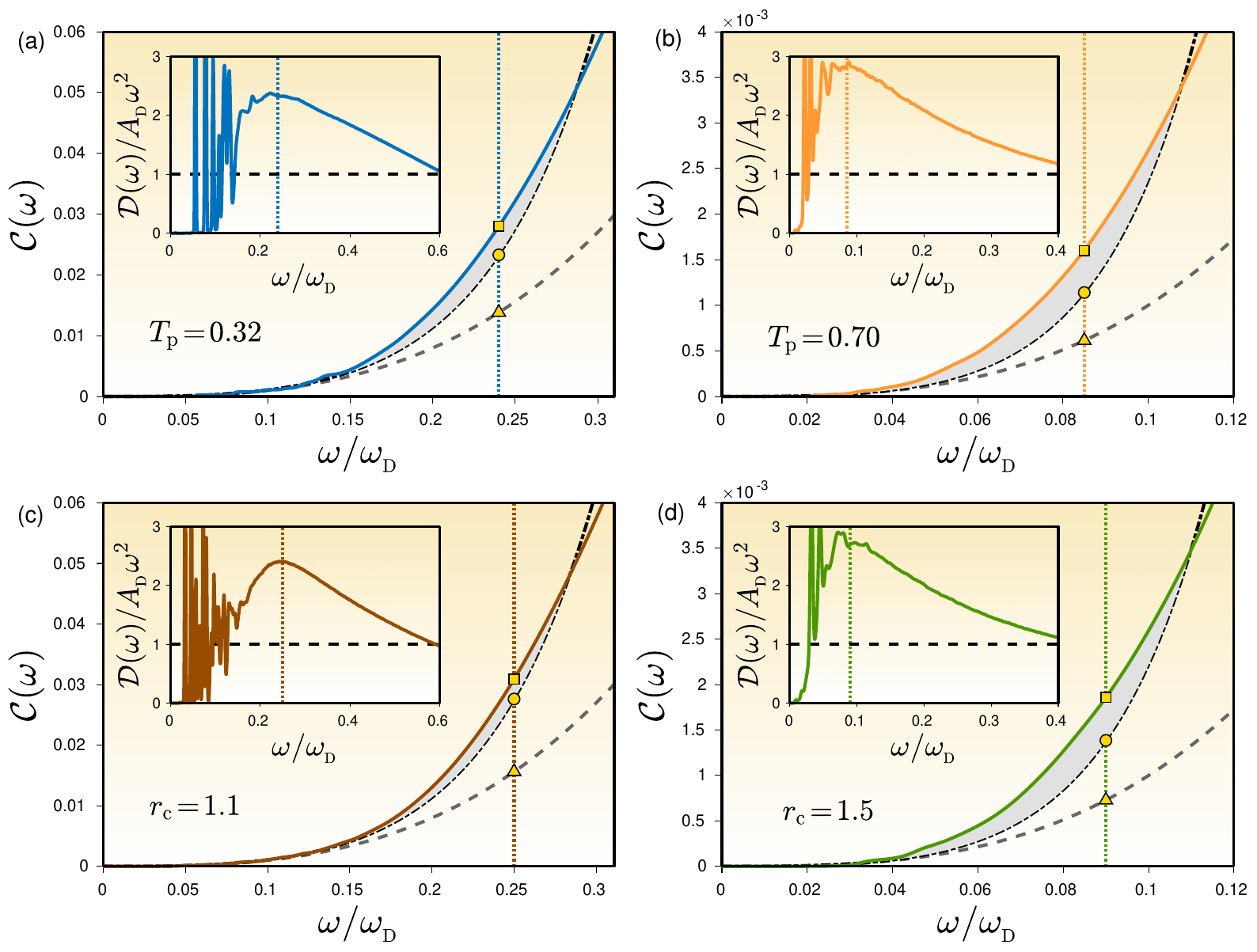}
  \caption{\footnotesize {\bf The different contributions to the boson peak}. (a) The cumulative VDoS ${\cal C}(\omega)\!=\!\int_0^\omega\!{\cal D}(\omega')d\omega'$ (solid line)
  for the $T_{\rm p}\!=\!0.32$ glass analyzed in Fig.~\ref{fig:length}b (the most deeply supercooled one). The vertical dashed line marks the BP frequency $\omega_{_{\rm BP}}$, which is determined using the reduced VDoS shown in the inset. Superposed are Debye's cumulative VDoS ${\cal C}_{_{\rm D}}\!(\omega)$ (dashed curve) and the cumulative phononic VDoS ${\cal C}_{\rm ph}(\omega)$ (dashed-dotted curve). The shaded area provides an estimate of the number of non-phononic vibrations, see text. The ratio ${\cal R}(\omega_{_{\rm BP}})$ of the distance between the square and the circle and the distance between the square and the triangle, corresponding to Eq.~\eqref{eq:ratio}, quantifies the relative contributions of non-phononic vibrations and nonlinear phonon dispersion to the BP. For this glass, we find ${\cal R}(\omega_{_{\rm BP}})\!=\!0.33$. (b) The same as panel (b), but for the $T_{\rm p}\!=\!0.70$ glass (the least deeply supercooled one used). Here, ${\cal R}(\omega_{_{\rm BP}})\!=\!0.47$. (c) The same as panel (a), but for the $r_{\rm c}\!=\!1.1$ glass analyzed in Fig.~\ref{fig:length}c. Here, ${\cal R}(\omega_{_{\rm BP}})\!=\!0.22$. (d) The same as panel (a), but the $r_{\rm c}\!=\!1.5$ glass analyzed in Fig.~\ref{fig:length}c. Here, ${\cal R}(\omega_{_{\rm BP}})\!=\!0.42$. In all cases, covering a broad range of glasses, ${\cal R}(\omega_{_{\rm BP}})$ is a sizable fraction of unity.}
  \label{fig:BP}
\end{figure*}

\vspace{-0.5cm}
\section{T\lowercase{he different contributions to the boson peak}}

To quantitatively determine the relative contributions of $\omega(k)$ and non-phononic vibrations to the BP, we develop a procedure illustrated in Fig.~\ref{fig:BP}a. For a given glass, whose $\omega(k)$ and ${\cal D}_{\rm ph}(\omega)$ have been computed as detailed above, we plot the reduced VDoS (inset) and determine the BP frequency $\omega_{_{\rm BP}}$. In the main panel, we plot the cumulative VDoS ${\cal C}(\omega)\=\int_0^\omega\!{\cal D}(\omega')d\omega'$ (solid line), and superpose on it Debye's cumulative VDoS ${\cal C}_{_{\rm D}}\!(\omega)$ (dashed line) and the cumulative phononic VDoS ${\cal C}_{\rm ph}(\omega)$ (dashed-dotted line), obtained by integrating Eq.~\eqref{eq:phonon_VDoS}. This procedure allows to extract several key quantities.

In particular, we computed the ratio
\begin{equation}
{\cal R}(\omega_{_{\rm BP}}) \equiv \frac{\displaystyle{\cal C}(\omega_{_{\rm BP}})-{\cal C}_{\rm ph}(\omega_{_{\rm BP}})}{\displaystyle{\cal C}(\omega_{_{\rm BP}})-{\cal C}_{_{\rm D}\!}(\omega_{_{\rm BP}})} \le 1
\label{eq:ratio}
\end{equation}
between the total number of non-phononic vibrations ${\cal C}_{_{\rm G}\!}(\omega_{_{\rm BP}})\={\cal C}(\omega_{_{\rm BP}})-{\cal C}_{\rm ph}(\omega_{_{\rm BP}})$, where ${\cal C}_{_{\rm G}\!}(\omega)$ is the cumulative non-phononic VDoS obtained by integrating ${\cal D}_{\!_{\rm G}\!}(\omega)$, and the total excess of vibrations on top of Debye's prediction ${\cal C}(\omega_{_{\rm BP}})-{\cal C}_{_{\rm D}\!}(\omega_{_{\rm BP}})$ up to the BP frequency $\omega_{_{\rm BP}}$. This procedure is performed for our smallest/largest $T_{\rm p}$ glass in Fig.~\ref{fig:BP}a/b, respectively, and for our smallest/largest $r_{\rm c}$ glass in Fig.~\ref{fig:BP}c/d, respectively. The ${\cal R}(\omega_{_{\rm BP}})$ values are reported in the caption and show that the two pieces of physics, i.e., non-phononic vibrations and nonlinear phonon dispersion, significantly contribute to the BP over a broad range of degrees of glassy disorder, including those characteristic of laboratory glasses. These findings cast doubt on suggestions that the BP is populated exclusively by softened phonons~\cite{Monaco_prl_2011,chumakov2016relation,nature_physics_boson_peak_2025}.

\vspace{-0.5cm}
\section{C\lowercase{onclusions}}

We developed a comprehensive and quantitative understanding of basic phenomena in disordered solids and their interrelations, including wave dispersion, wave attenuation and non-Debye vibrational spectra, using scaling theory and various computational tools. Specifically, by studying three classes of computer disordered solids, we showed that the nonlinearity in the phonon dispersion relation $\omega(k)$ in these systems is controlled by a disorder-induced lengthscale $\xi\!\gg\!a_{_0}$, where $a_{_0}$ is an average interparticle distance. The mesoscopic lengthscale $\xi$, which does not exist in pure crystals, has also been shown to control the Rayleigh scattering scaling of the wave attenuation rate $\Gamma(k)\!\sim\!\xi^2 k^4$ for $k\xi\!\ll\!1$. Our findings regarding the leading-order nonlinearity in $\omega(k)$ and the asymptotic behavior of $\Gamma(k)$ are scaling-wise consistent with the continuum-based HET~\cite{schirmacher2006thermal,Schirmacher_prl_2007,Schirmacher_2013_boson_peak}. At the same time, our findings also point to some serious difficulties in recent work, see Appendix~\ref{sec:NatPhy2025}.

Our robust computation of $\omega(k)$ in disordered solids, controlled by $\xi$, allowed us to quantitatively predict the two distinct physical contributions to low-frequency deviations of the VDoS of disordered solids ${\cal D}(\omega)$ from Debye's prediction ${\cal D}_{_{\rm D}}\!(\omega)\=A_{_{\rm D}}\omega^2$. The first is phononic in nature, emerging from nonlinearity in $\omega(k)$, leading to ${\cal D}_{\rm ph}(\omega)$ that deviate from ${\cal D}_{_{\rm D}}(\omega)$ with increasing $\omega$ due to $\xi$. The second is non-phononic in nature, corresponding to non-phononic vibrations in glasses that follow their own VDoS ${\cal D}_{\!_{\rm G}\!}(\omega)$, which cannot be captured by continuum-based approaches. ${\cal D}_{\!_{\rm G}\!}(\omega)$ is known to feature a power-law tail $\sim\!\omega^4$ and to increase significantly at higher frequencies.

We showed that the phononic and non-phononic contributions to the onset of a non-Debye behavior in glasses, i.e., the leading $\sim\!\omega^4$ terms in excess of ${\cal D}_{_{\rm D}}\!(\omega)\=A_{_{\rm D}}\omega^2$, can feature widely different ratios depending on the degree of disorder of glasses, where the phononic contribution is expected to dominate for laboratory glasses formed by a quenched. Finally, and crucially, we quantified the fraction ${\cal R}(\omega_{_{\rm BP}})$ of non-phononic vibrations out of the total excess modes on top of Debye's prediction at the BP frequency $\omega_{_{\rm BP}}$. We found that ${\cal R}(\omega_{_{\rm BP}})$ is a sizable fraction of unity over a broad range of physical conditions, demonstrating that both phononic and non-phononic vibrations significantly contribute to the BP.

Our results and analysis framework can be directly applied to experiments, offering testable predictions. Specifically, once $\omega(k)$, $A_{_{\rm D}}$ and ${\cal D}(\omega)$ are measured, the basic quantity ${\cal R}(\omega_{_{\rm BP}})$ can be obtained and also compared to the theoretical expectations. From a broader perspective, the relevance and importance of the BP modes composition for glassy phenomena such as transport and plasticity should be explored.

\appendix

\section{Computer models and parameters extraction}
\label{sec:models}
\label{sec:appendix_computer_models}

\begin{figure}[ht!]
  \includegraphics[width = 0.49\textwidth]{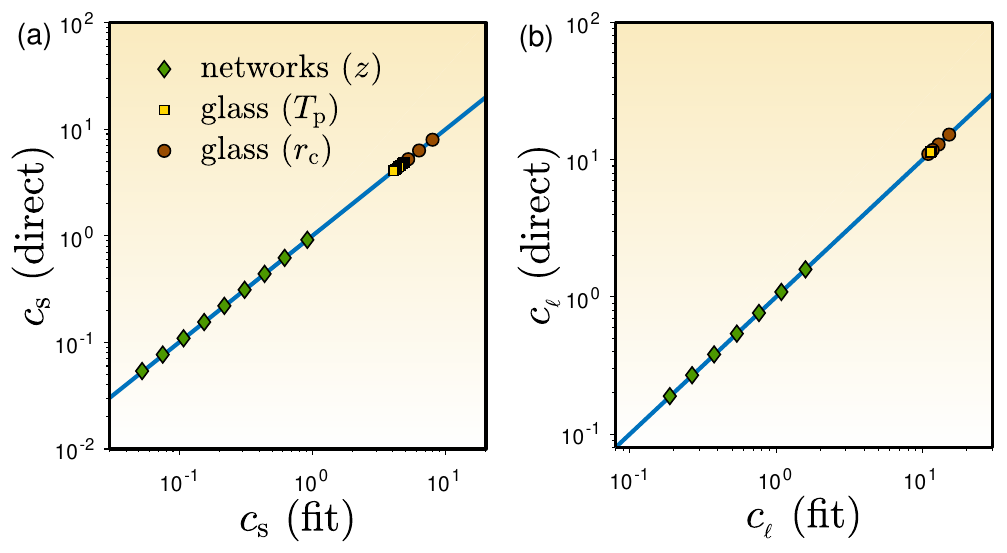}
  \caption{\footnotesize \textbf{Validation of wave-speed fits.} The shear ($c_{\rm s}$, panel (a)) and longitudinal ($c_{_\ell}$, panel (b)) wave speeds as obtained from the cubic dispersion fits (see Tables~\ref{table:networks},~\ref{table:PIPL} and~\ref{table:sticky_spheres}), plotted against direct computations obtained via the elastic moduli of our model solids. An excellent agreement is observed.}
  \label{fig:wave_speed_validation}
\end{figure}

\begin{figure*}[ht!]
  \includegraphics[width = 0.82\textwidth]{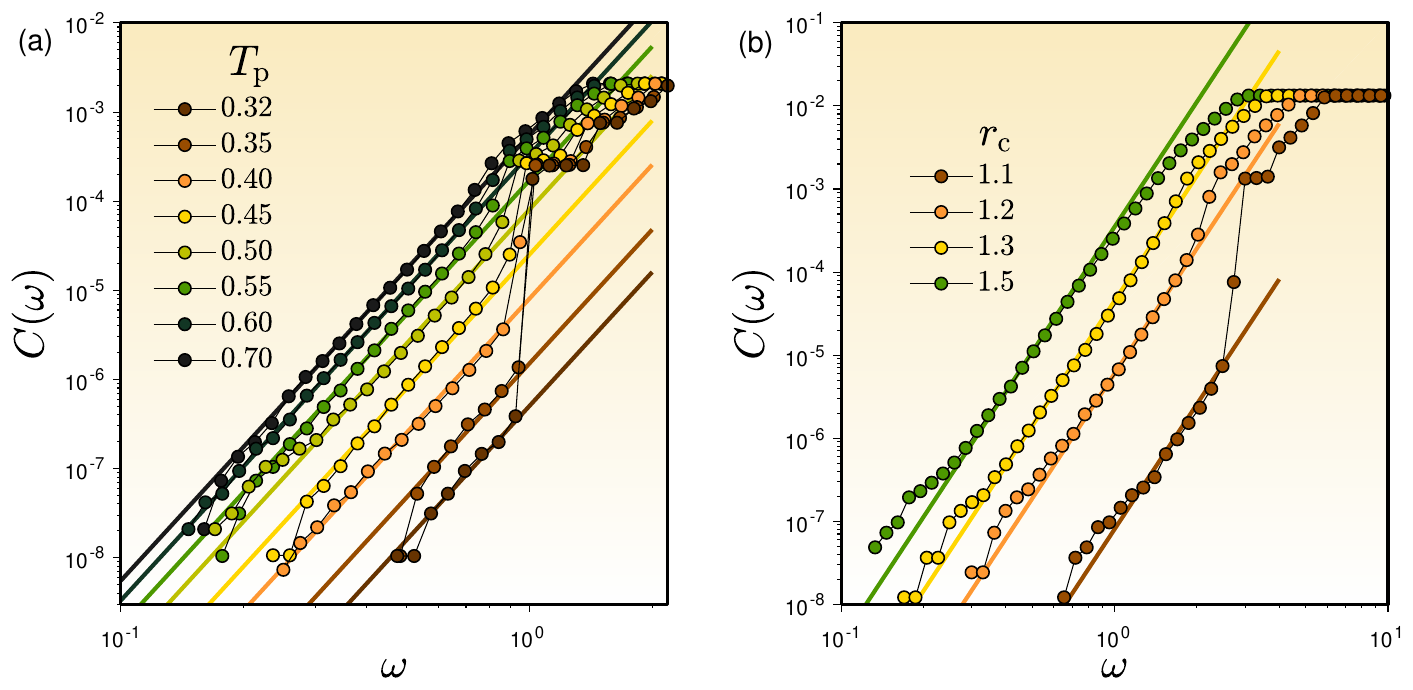}
  \caption{\footnotesize \textbf{Extraction of the non-phononic VDoS prefactor $A_{\rm g}$ in computer glasses.} Low-frequency tails of (a) the polydisperse (PIPL) glass with $N\!=\!16,000$, and (b) the sticky-spheres glass with $N\!=\!3000$. The extracted values of $A_{\rm g}$ can be found in Tables~\ref{table:polyipl} and~\ref{table:sticky_spheres} for the two models, respectively.}
  \label{fig:extract_Ag}
\end{figure*}

In this work, we employ four models of 3D disordered solids, belonging to three different classes each characterized by a control parameter, and one model of a crystalline solid, as described in the main text and in more detail below. For all models, the four physical parameters $c_{\rm s}$, $c_{_{\ell}}$, $\xi_{\rm s}$ and $\xi_{_{\ell}}$ were obtained via fits of the measured wave dispersion to $\omega_{{\rm s},\ell}(k)\!\simeq\!c_{{\rm s},\ell}k\!-\!c_{{\rm s},\ell}\Upsilon\xi_{{\rm s},\ell}^2k^3$, i.e., the leading-order nonlinear expansion of Eq.~\eqref{eq:nonlinear_dispersion}, with $\Upsilon\!=1/96$. In Fig.~\ref{fig:wave_speed_validation}, we compare the wave speeds as obtained via the dispersion fits with the wave speeds obtained by direct measurements of elastic moduli. In addition, for each glass model and each value of its respective control parameter, the prefactor $A_{\rm g}$ of Eq.~\eqref{eq:QLMs_VDoS_tail} was obtained by fitting the low-frequency tail of the cumulative VDoS ${\cal C}(\omega)\!\equiv\!\int_0^\omega{\cal D}(\omega')d\omega'\!=\!(A_{\rm g}/5)\omega^5$ of small glassy samples, see Fig.~\ref{fig:extract_Ag}. The extracted parameters are reported below, along with the values of the disorder quantifier $\chi$, which is discussed in the main text and is formally defined in Appendix~\ref{sec:chi}.

\subsection{Lennard-Jones FCC crystal}
\vspace{-0.2cm}
We place particles interacting via a standard Lennard-Jones potential in an FCC crystal structure. The potential is smoothed at the interaction cutoff as described in~\cite{boring_paper}. We set the number density at $N/V\!=\!1.0467$ (in simulation units), resulting in a vanishing hydrostatic pressure at zero temperature. We employ a system of $N\=16,384,000$ particles for the VDoS calculation in Fig.~\ref{fig:onset}a of the main text. By fitting the low-frequency tail to $3\omega^2\!/\omega_{_{\rm D}}^3$, we find $\omega_{_{\rm D}}\!\approx\!23.54$ (in simulational units).

\subsection{Disordered Hookean spring networks}
\vspace{-0.2cm}

We employ disordered networks of relaxed Hookean springs that are derived from binary inverse-power-law (BIPL) glasses, to be described below. Springs are diluted from the initial dense interaction network to reach a target coordination $z$ following the algorithm described in~\cite{anomalous_elasticity_soft_matter_2023}, which maintains low coordination fluctuations. We varied $N$ between $10^3$ and $4\!\times\!10^6$, and the coordination $z$ between $6.04$ to $16.24$. In Table~\ref{table:networks}, we provide all the relevant parameters for this model. We note that we could not reliably extract $c_{_\ell}$ and $\xi_{_\ell}$ for $z\!=\!6.08$ and $z\!=\!6.04$ with the system sizes employed.

\begin{table}[h]
    \centering
    \caption{Extracted parameters of the disordered networks}
    \label{table:networks}
    \begin{tabular}{| c | c c c c c|}
        \hline
        $z$ & $c_{\rm s}$ & $c_{_{\ell}}$ & $\xi_{\rm s}$ & $\xi_{_{\ell}}$ & $\chi$\\ \hline
        16.24 & 0.92 & 1.59 & 2.50 & 3.79 & 0.23 \\ \hline
        11.12 & 0.62 & 1.09 & 3.02 & 4.75 & 0.57\\ \hline
        8.56 & 0.44 & 0.76 & 4.38 & 7.46 & 0.99\\ \hline
        7.28 & 0.31 & 0.54 & 6.35 & 10.73 & 1.60\\ \hline
        6.64 & 0.22 & 0.38 & 9.03 & 14.53 & 2.33\\ \hline
        6.32 & 0.15 & 0.27 & 13.15 & 21.5 & 3.60\\ \hline
        6.16 & 0.11 & 0.19 & 17.8 & 31.3 & 5.11\\ \hline
        6.08 & 0.075 & n/a & 24.0 & n/a & 7.16 \\ \hline
        6.04 & 0.053 & n/a & 34.6 & n/a & 10.3 \\ \hline
    \end{tabular}
\end{table}

\subsection{Polydisperse inverse-power-law (PIPL) glasses}
\vspace{-0.2cm}
The polydisperse inverse-power-law (PIPL) glass former employs a purely repulsive, $\sim\!1/r^{10}$ pairwise potential, where the effective size parameters of the particles are drawn from a distribution that optimizes the applicability of the swap-Monte-Carlo algorithm for very deep supercooling~\cite{LB_swap_prx}. Details of this model can be found in~\cite{boring_paper}. The number density is set to $N/V\!=\!0.58$, and we employ system sizes ranging from $N\!=\!16,000$ to $N\!=\!256,000$. In Table~\ref{table:PIPL}, we provide all the relevant parameters for this model.

\begin{table}[h]
    \centering
    \caption{Extracted parameters of the PIPL glasses}\label{table:PIPL}
    \begin{tabular}{| c | c c c c c c|}
        \hline
        $T_{\rm p}$ & $c_{\rm s}$ & $c_{_{\ell}}$ & $\xi_{\rm s}$ & $\xi_{_{\ell}}$ & $A_{\rm g}$ & $\chi$\\ \hline
        0.32 & 4.903 & 11.565 & 4.27 & 6.20 & 2.5e-06  & 0.95 \\ \hline
        0.35 & 4.835 & 11.557 &    4.60 & 6.57 & 7.5e-06 & 1.0 \\ \hline
        0.40 & 4.72 & 11.54 & 5.36 & 7.40 & 4.0e-05 & 1.2  \\ \hline
        0.45 & 4.58 & 11.50 & 6.27  & 8.49 & 1.2e-04 & 1.45 \\ \hline
        0.50 & 4.45 & 11.476 & 8.31
        & 10.87 & 4.0e-04 & 1.95 \\ \hline
        0.55 & 4.305 & 11.425 & 9.60 & 12.20 & 8.5e-04 & 2.4 \\ \hline
        0.60 & 4.184 &  11.37 & 11.92 & 14.37 & 1.6e-03 & 2.9 \\ \hline
        0.70 & 4.053 & 11.35  & 12.96 & 15.74 &  2.7e-03  & 3.4 \\ \hline
    \end{tabular}
\label{table:polyipl}
\end{table}

\subsection{Binary inverse-power-law (BIPL) glasses}
\begin{figure}[ht!]
  \includegraphics[width = 0.49\textwidth]{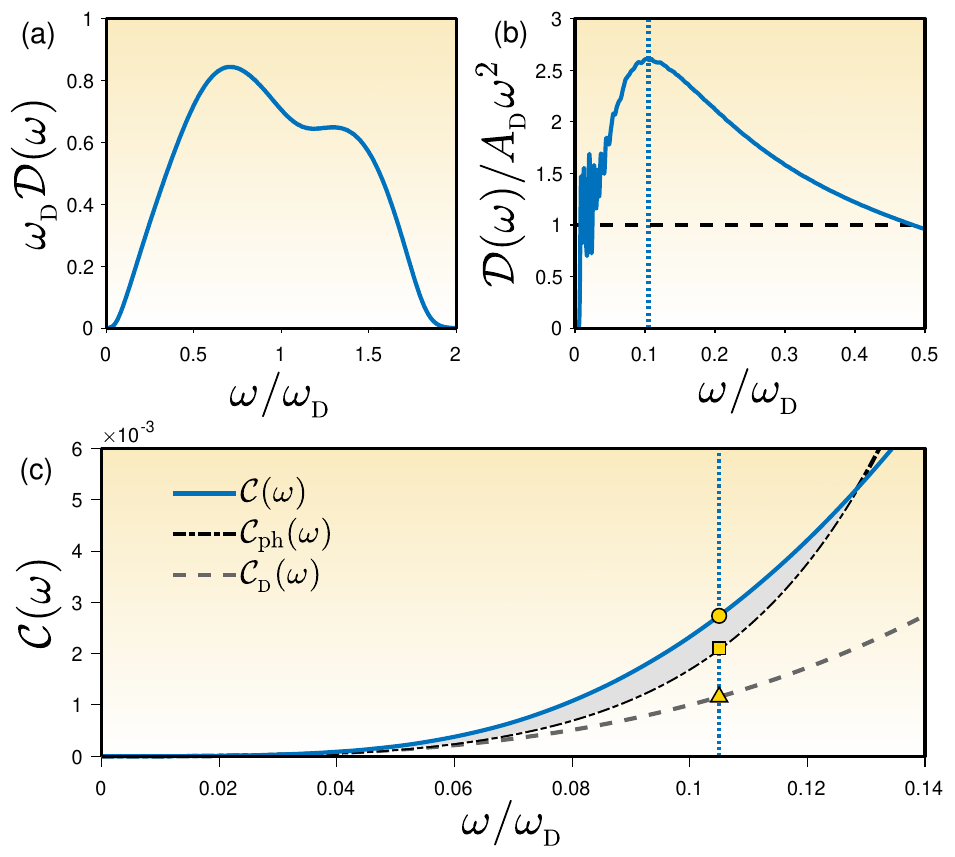}
  \caption{\footnotesize \textbf{VDoS of the hyperquenched binary BIPL glass.} Here, we follow the analysis presented in Fig.~\ref{fig:BP}. (a) The VDoS ${\cal D}(\omega)$ calculated using $N\!=\!4,000,000$ glasses. (b) The reduced VDoS ${\cal D}(\omega)/A_{_{\rm D}}\omega^2$, featuring a boson peak (marked by the vertical dotted line). (c) Cumulative VDoS ${\cal C}(\omega)$ (blue line), cumulative phononic VDoS ${\cal C}_{\rm ph}(\omega)$ (dash-dotted line), and cumulative Debye VDoS ${\cal C}_{_{\rm D}\!}(\omega)$ (dashed line). We find ${\cal R}(\omega_{_{\rm BP}})\!\simeq\!0.4$.}
  \label{fig:3dipl_cumulative}
\end{figure}

\vspace{-0.2cm}
\label{sec:appendix_BIPL}
The binary inverse-power-law (BIPL) is a 50:50 binary mixture of `large' and `small' particles interacting via the same purely repulsive $\sim\! 1/r^{10}$ pairwise potential as employed in the PIPL model. Details about the model can be found in~\cite{modes_prl_2018}. Glasses are quenched from high-temperature equilibrium states at a rate of $10^{-3}$, and the number density is fixed to $N/V\!=\!0.82$ (both in simulational units). We employ systems of $N\!=\!256,000$ particles for measuring the wave dispersion $\omega(k)$, and the VDoS was calculated using $N\!=\!4000$ (see Fig.~\ref{fig:onset}c of the main text) and $N\!=\!4\times10^6$ (see inset of Fig.~\ref{fig:onset}c and Fig.~\ref{fig:onset}d of the main text). In Table~\ref{table:BIPL}, we provide all the relevant parameters for this
model. In addition, for completeness, Fig.~\ref{fig:3dipl_cumulative} shows the same analysis of Fig.~\ref{fig:BP} of the main text applied to this model (here $N\!=\!4 \times 10^6$), where we find ${\cal R}(\omega_{_{\rm BP}})\!\simeq\!0.4$.

\begin{table}[h]
    \centering
    \caption{Extracted parameters of the BIPL glasses}\label{table:BIPL}
    \begin{tabular}{| c c c c c c|}
        \hline
         $c_{\rm s}$ & $c_{_{\ell}}$ & $\xi_{\rm s}$ & $\xi_{_{\ell}}$ & $A_{\rm g}$ & $\chi$ \\ \hline
         4.1685 & 11.33 & 9.2 & 11.9 & 5.5e-4 & 2.7
     \\ \hline
    \end{tabular}
\label{table:binary_ipl}
\end{table}

\subsection{Sticky-sphere binary glasses}
\vspace{-0.2cm}
The so-called sticky-sphere glass is a 50:50 binary mixture of `large' and `small' particles interacting via a Lennard-Jones-like pairwise potential with a tunable attractive branch controlled by the dimensionless parameter $r_{\rm c}$: a larger $r_{\rm c}$ results in a weaker attraction accompanied by a longer interaction range. All details about the interaction potential (introduced first in~\cite{itamar_sticky_spheres_potential_pre_2011}) can be found in~\cite{sticky_spheres1_karina_pre2021}, along with a comprehensive study regarding how $r_{\rm c}$ affects the resulting glasses' elastic properties. We studied this model with $r_{\rm c}\!=\!1.1,1.2,1.3,1.5$. We set the number density at $N/V\!=\!0.60$, and set the number of particles at $N\!=\!81,000$ for the dispersion and VDoS calculations, and at $N\!=\!3000$ for extracting the prefactor $A_{\rm g}$ of the non-phononic VDoS low-frequency tail. Glasses were quenched to $T\!=\!0$ from high-temperature parent equilibrium states, specifically $T_{\rm p}\!=\!6.0,4.0,4.0,2.6$ for the aforementioned values of $r_{\rm c}$, respectively. In Table~\ref{table:sticky_spheres}, we provide all the relevant parameters for this model.

\begin{table}[h]
    \centering
    \caption{Extracted parameters of the sticky-sphere glasses}
    \begin{tabular}{| c | c c c c c c|}
        \hline
        $r_{\rm c}$ & $c_{\rm s}$ & $c_{_{\ell}}$ & $\xi_{\rm s}$ & $\xi_{_{\ell}}$ & $A_{\rm g}$ & $\chi$ \\ \hline
        1.1 & 7.953 & 15.24  & 4.27  & 6.57  & 4.0e-07 & 1.0 \\ \hline
        1.2 & 6.305 & 12.915 & 6.50  & 9.34  & 3.0e-05 & 1.85 \\ \hline
        1.3 & 5.223 & 11.77  & 8.93  & 11.72 & 2.2e-04 & 2.6 \\ \hline
        1.5 & 4.440 & 10.97  & 12.39 & 15.18 & 1.7e-03 & 3.6 \\ \hline
    \end{tabular}
\label{table:sticky_spheres}
\end{table}

\begin{figure}[ht!]
  \includegraphics[width = 0.45\textwidth]{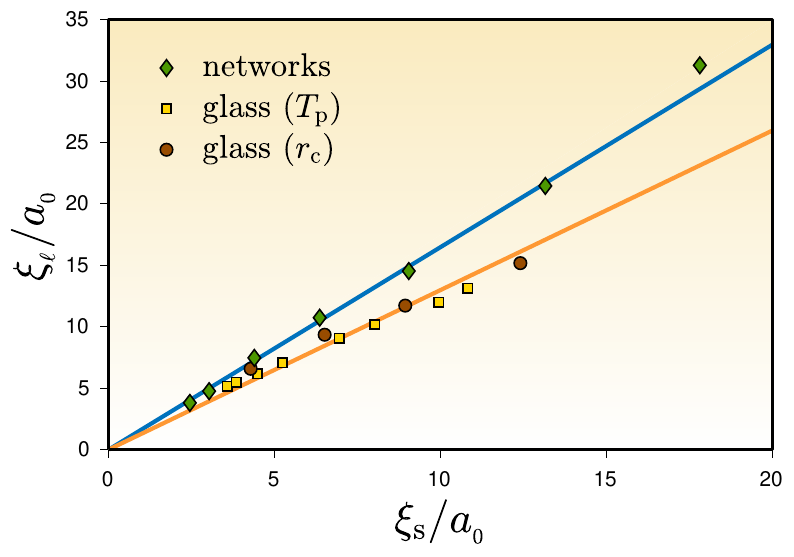}
\caption{\footnotesize \textbf{Proportionality of the shear and longitudinal disorder-induced dispersion lengthscales.} The longitudinal dispersion length $\xi_{_\ell}$ is plotted against the shear dispersion length $\xi_{\rm s}$, for our networks, the PIPL model and the sticky spheres model. We find $\xi_{_\ell}\!\sim\!\xi_{\rm s}$ to a good approximation. The blue and orange lines have slopes 1.65 and 1.3, respectively.}
\label{fig:compare_xis_xil}
\end{figure}

The above results allow us to establish the approximate relation $\xi_{\rm s}\!\sim\!\xi_{_\ell}$, as stated in the main text and explicitly demonstrated in Fig.~\ref{fig:compare_xis_xil}.

\section{Methods}
\label{sec:methods}
\vspace{-0.2cm}

The results of Appendix~\ref{sec:models} involved the dispersion relation $\omega(k)$ and the VDoS ${\cal D}(\omega)$. Here, we explain how these basic physical quantities are observed. We also define the disorder quantifier $\chi$, whose values are reported in Appendix~\ref{sec:models}.

\begin{figure*}[ht!]
  \includegraphics[width = 0.85\textwidth]{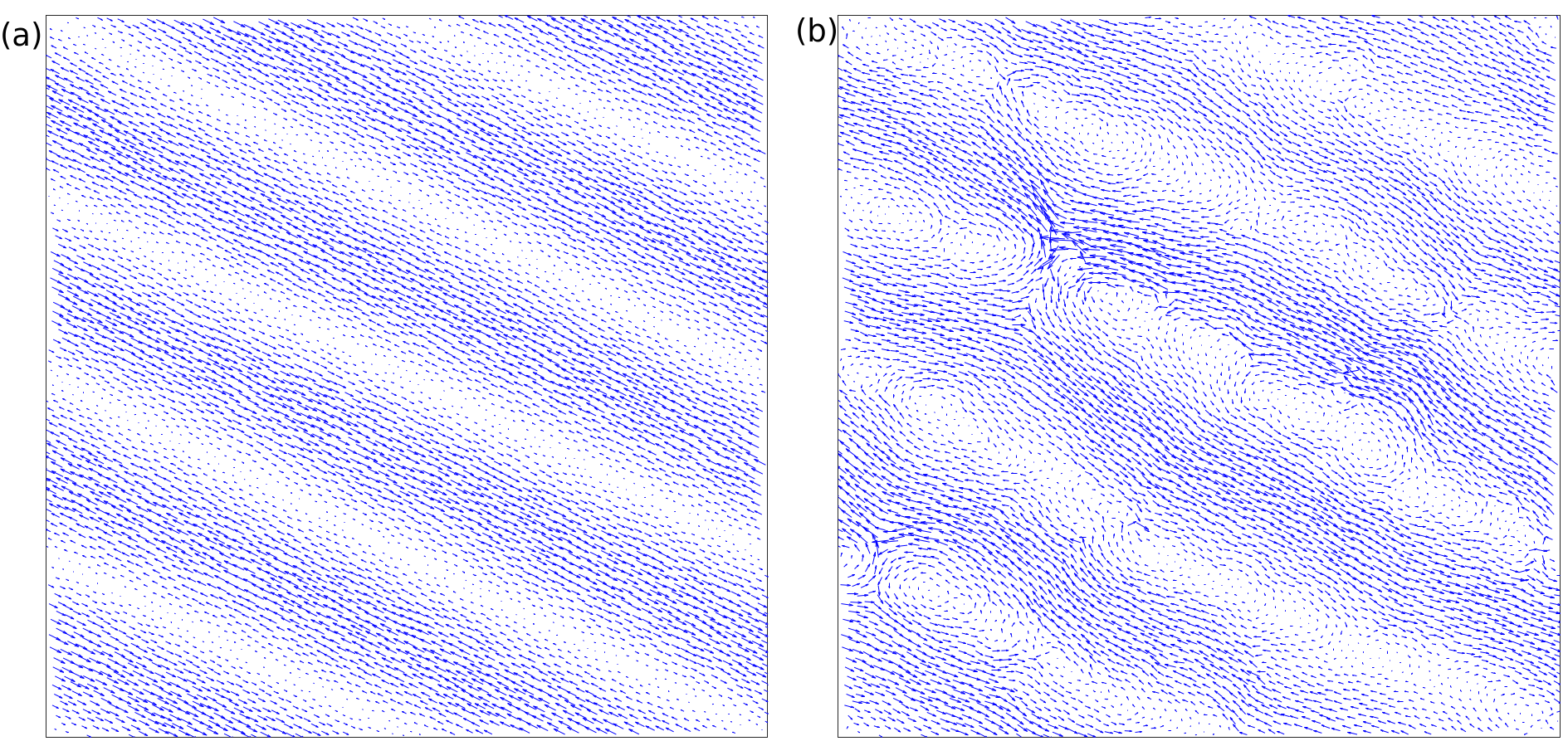}
  \caption{\footnotesize \textbf{Illustration of the `imposed-wave' method.} (a) An example of the imposed wave ${\bm f}(\kv)$ of Eq.~\eqref{eq:imposed_wave} in a 2D glass of $N\!=\!16,000$ particles. (b) The linear response $\uv(\kv)$ to the imposed wave, see Eq.~\eqref{eq:linear_response} and text for discussion.}
  \label{fig:fake_phonon_example}
\end{figure*}

\subsection{Elastic wave dispersion measurements:\\ The imposed-wave method}
\vspace{-0.2cm}
\label{sec:appendix_imposed_wave}

In order to compute the phonon dispersion relation $\omega(k)$ in disordered solids, we developed and applied the so-called `imposed-wave method', described below. As explained in the main text, the main challenge in computing $\omega(k)$ in the presence of disorder is the unique identification of the frequency $\omega$ of a phonon of wavenumber $k$.

Our idea is to impose on each particle/node in a disordered solid a force vector that varies in space like a wave/phonon (either shear/transverse or longitudinal/sound) with an admissible wavevector ${\bm k}$ and to compute the resulting displacement in the linear response approximation. The resulting displacement, while not being an exact eigenfunction of the solid's Hessian $\calBold{H}$, can be associated with a frequency in complete analogy to exact eigenfunctions. The assigned frequency to a wavenumber $k\=|{\bm k}|$ can be averaged over directions and quenched disorder. Similar procedures were put forward in~\cite{liu_soft_matter_2013,brian_prl_2017}.

To formalize these ideas, we define imposed force waves as
\begin{equation}
    {\bm f}_i^{({\rm s},\ell)}(\kv) = \hat{\kv}^{({\rm s},\ell)}\,e^{i\kv\cdot\rv_i}-N^{-1}\sum_j\hat{\kv}^{({\rm s},\ell)}\,e^{i\kv\cdot\rv_j}\,,
\label{eq:imposed_wave}
\end{equation}
where the index $i$ labels the $i^{\mbox{\tiny th}}$ particle (in glasses) or node (in networks), $\kv$ is an admissible wavevector and $\hat{\kv}^{({\rm s},\ell)}$ are polarization unit vectors corresponding to shear and longitudinal waves/phonons, respectively. The polarization of longitudinal phonons reads $\hat{\kv}^{({\ell})}\!=\!\kv/|\kv|$ and hence that of shear phonons is $\hat{\kv}^{({\rm s})}\!\cdot\!\hat{\kv}^{(\ell)}\!=\!0$ (of course there are two such shear polarizations). The admissible wavevectors are $\kv\!=\!2\pi\nv/L$, where $L$ the system's linear size, and $\nv\!=\!(n_x,n_y,n_z)$ is a vector of integers that are selected by solving the integer sum of squares problem in 3D~\cite{phonon_widths}.

The imposed wave ${\bm f}(\kv)$ in Eq.~\eqref{eq:imposed_wave} includes two terms. The first term is just a pure wave/phonon. Yet, since the actual particle/node positions $\{\rv_i\}$ feature positional disorder, the first term also includes disorder-induced projections on collective translations of the system's center of mass, which should be removed. This is achieved through the second term. For the sake of brevity, we omit hereafter the superscripts and subscripts $({\rm s},\ell)$ from all subsequent definitions, but they are implicitly implied. An example of ${\bm f}(\kv)$ in a 2D glass is shown in Fig.~\ref{fig:fake_phonon_example}a (our analysis is performed in 3D, but for visual clarity and illustration purposes we present results in 2D).

The linear response to ${\bm f}(\kv)$ is given by
\begin{equation}
    \uv(\kv) = \calBold{H}^{-1}\cdot{\bm f}(\kv)\,.
\label{eq:linear_response}
\end{equation}
where $\calBold{H}^{-1}$ is the inverse of the Hessian. An example of $\uv(\kv)$ is shown in Fig.~\ref{fig:fake_phonon_example}b, demonstrating that its spatial structure closely resembles a phononic vibrational mode. If $\uv(\kv)$ was an exact eigenfunction of $\calBold{H}$, its eigenvalue would be the associated frequency squared $\omega^2$. In complete analogy, we associate with $\uv(\kv)$ the frequency squared $[\omega(\kv)]^2$ according to
\begin{equation}
    [\omega(\kv)]^2 = \frac{\uv(\kv)\cdot\calBold{H}\cdot\uv(\kv)}{\uv(\kv)\cdot\uv(\kv)} = \frac{{\bm f}(\kv)\cdot\calBold{H}^{-1}\cdot{\bm f}(\kv)}{{\bm f}(\kv)\cdot\calBold{H}^{-2}\cdot{\bm f}(\kv)}\,.
\label{eq:wave_freq}
\end{equation}
The frequency $\omega(k)$ (note that here the wavenumber $k\=|\kv|$ is the argument) is obtained by averaging $\omega(\kv)$ of Eq.~\eqref{eq:wave_freq} over various wavevectors $\kv$ that share the same magnitude $k\!=\!|\kv|$ and to improve convergence also over independent disordered solid realizations.

\begin{figure}[ht!]
  \includegraphics[width = 0.49\textwidth]{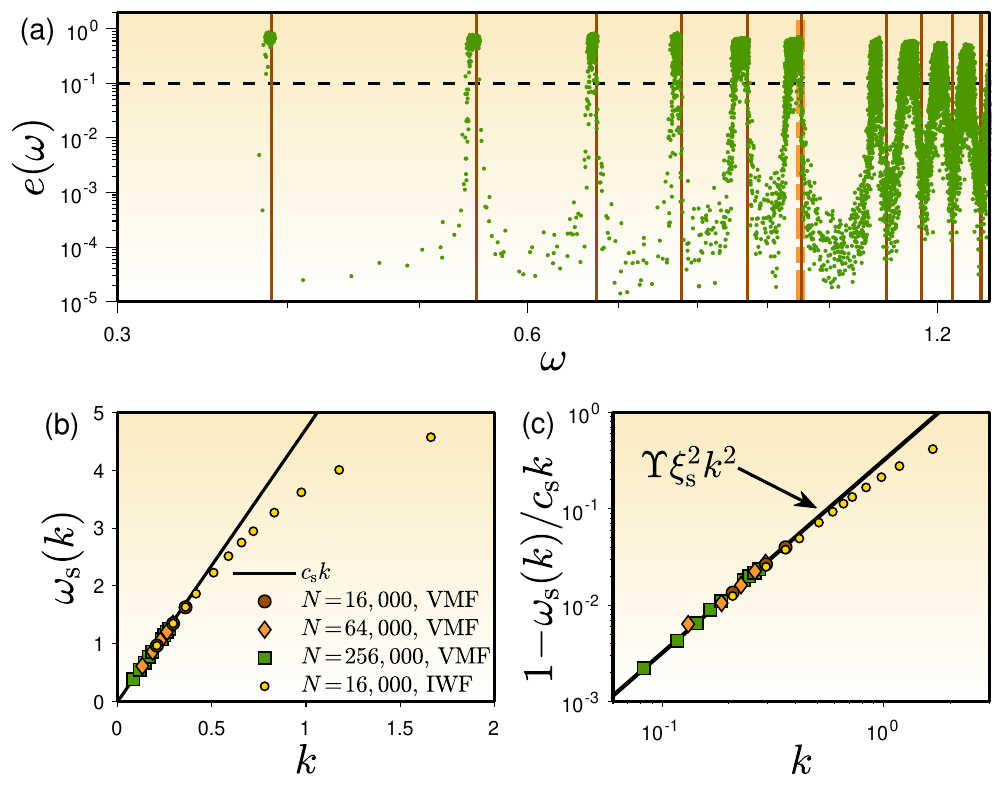}
  \caption{\footnotesize \textbf{Validation of the `imposed-wave' method.} (a) The participation ratio $e(\omega)$ scatter-plotted against frequency $\omega$ of low-frequency vibrational modes in the PIPL model with $T_{\rm p}\!=\!0.40$ and $N\!=\!256,000$ particles. The vertical lines represent the estimations $c_{\rm s}k$ of phonon band frequencies assuming a linear dispersion. The dashed vertical line represents the lowest-frequency sound wave. The horizontal dashed line marks the chosen threshold for identifying phonons in each band, see text for additional details.}
  \label{fig:fake_phonons}
\end{figure}

In order to validate that the imposed-wave method produces the correct dispersion of elastic waves, we consider the PIPL glass model with parent temperature $T_{\rm p}\!=\!0.40$, which is moderately supercooled prior to glass formation. In Fig.~\ref{fig:fake_phonons}a, we scatter-plot the participation ratio $e(\omega)$ of vibrational modes vs.~their frequency $\omega$, for $N\!=\!256,000$. Over the range of frequencies shown, phonon frequencies are grouped into discrete bands; we mark the expected frequencies of  transverse phonons $c_{\rm s}k$ with vertical lines. The nonlinearity of phonon dispersion is apparent by the occurrence of phonon bands progressively below the expected linear-dispersion frequencies $c_{\rm s}k$ as one considers higher frequencies.

In order to obtain the true dispersion, we set a threshold $e_{\rm th}\!=\!0.1$ --- marked by the horizontal dashed line in Fig.~\ref{fig:fake_phonons}a ---, and take the average frequency of modes with $e(\omega)\!>\!e_{\rm th}$ within each phonon band. Since quasi-localized modes feature $e(\omega)\!\sim\!N^{-1}\!\ll\!e_{\rm th}$, their effect on the averages is minimal. The resulting averages $\omega_{\rm s}(k)$ are shown in Fig.~\ref{fig:fake_phonons}b (vibrational-modes frequencies are denoted as `VMF' in the figure legend), for various $N$. We superimpose the results of our imposed-wave frequencies (denoted as `IWF' in the figure legend) to find very good agreement with the direct estimations from the vibrational modes. Additionally, in Fig.~\ref{fig:fake_phonons}c we show that our estimations of $\xi_{\rm s}$ using VMF or IWF agree very well, and appear to be $N$-independent.

Apart from its simplicity, our approach has two main advantages: first, we do not rely on any modeling that may bias the resulting dispersion curve $\omega(k)$. For instance, in~\cite{tanaka_2d_modes_2022} different models are used to fit the transverse and longitudinal dynamic structure factors $S_{_{{\rm T},{\rm L}\!}}(k,\omega)$. Second, our calculation provides very accurate estimations of the dispersion curve, allowing us to reliably extract the dispersion length $\xi$.

\subsection{The Kernel Polynomial Method (KPM)}
\begin{figure}[ht!]
  \includegraphics[width = 0.45\textwidth]{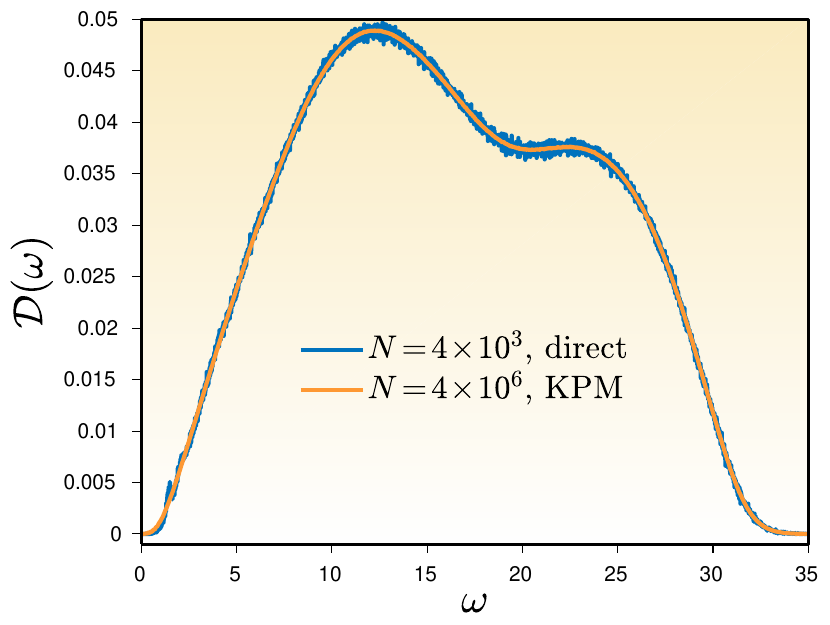}
  \caption{\footnotesize \textbf{Benchmark of our KPM calculations.} The VDoS ${\cal D}(\omega)$ of the BIPL glass calculated with the KPM (orange curve) for systems of $N\!=\!4\!\times\!10^6$ particles, and by direct diagonalization for systems of $N\!=\!4\!\times\!10^3$ particles. Note that the VDoS obtained using a direct diagonalization in the smaller systems features discrete phonon bands at the low-frequency end of the spectrum.}
  \label{fig:kpm_validation}
\end{figure}

\vspace{-0.2cm}
In order to obtain vibrational spectra ${\cal D}(\omega)$ of very large computer glasses, needed for quantitatively resolving the low-frequency vibrational properties, we employ the Kernel Polynomial Method (KPM) as described, e.g., in~\cite{Tanguy_2016}. The method requires 3 input parameters: $\omega_{\rm max}$ is a bound on the maximal vibrational frequency in the system, $K$ is the truncation degree that determines the frequency resolution $\delta\omega\!\sim\!K^{-1}$, and $R$ denotes the number of independent random realizations of the algorithm. We chose $K\=3000$ and $R\!=10$ for all calculations. In Fig.~\ref{fig:kpm_validation}, we benchmark our KPM calculations against a direct diagonalization of the Hessian matrices of smaller glass samples.

\subsection{The disorder quantifier $\chi$}
\label{sec:chi}
\vspace{-0.2cm}

The broadly applicable disorder quantifier $\chi$ has been studied extensively in previous work; it was shown to control wave attenuation rates in disordered solids~\cite{jcp_letter_scattering_2021}, to control phonon-band widths in the vibrational spectra of finite-size computer glasses~\cite{phonon_widths}, and to follow scaling laws in the vicinity of the unjamming point~\cite{julia_chi_2024}. Furthermore, it was shown~\cite{phonon_widths2} to be closely related to the `disorder parameter' $\gamma$ of HET~\cite{schirmacher2006thermal,Schirmacher_prl_2007,Schirmacher_2013_boson_peak}.

Given an ensemble of disordered solids composed of $N$ particles/nodes, formed by some protocol, and each member featuring a shear modulus $\mu$, the disorder quantifier $\chi$ is defined as
\begin{equation}\label{eq:chi_definition}
    \chi = \sqrt{N}\frac{\sqrt{\langle(\Delta \mu)^2\rangle}}{\langle \mu \rangle}\,,
\end{equation}
where $\Delta \mu\!\equiv\!\mu\!-\!\langle \mu \rangle\!\sim\!1/\sqrt{N}$ and the triangular brackets denote an ensemble average. When measuring $\chi$ in computer glasses, it is advantageous to employ an outlier-removal strategy (see, e.g.,~\cite{sticky_spheres1_karina_pre2021,phonon_widths2}), since the probability distribution of $\mu$ features a fat tail~\cite{jcp_letter_scattering_2021} leading to noisy results. In this work, we used the measurements of $\chi$ for the PIPL model reported in~\cite{phonon_widths2} (see Table~\ref{table:PIPL}), and for the sticky spheres model reported in~\cite{sticky_spheres1_karina_pre2021} (see Table~\ref{table:sticky_spheres}). We calculated $\chi$ for our disordered spring networks via Eq.~(\ref{eq:chi_definition}), and report the results in Table~\ref{table:networks}.

\section{Relating $\omega(k)$ to the phononic VDoS}
\vspace{-0.2cm}
\label{sec:appendix_A6}

Since the dispersion of each acoustic branch is an odd function of $k$, it follows a form
\begin{equation}
    \omega(k) \simeq ck - c\Upsilon\xi^2k^3\,,
\label{eq:cubic_dispersion_SI}
\end{equation}
where $\Upsilon\!=\!1/96$ (see main text) and $c$ is the relevant wave speed. The corresponding inverse $k(\omega)$ of the dispersion reads
\begin{equation}
    k(\omega)
  \simeq \frac{\omega}{c}
   + \frac{\Upsilon\xi^2}{c^3}\,\omega^3
   + \frac{3\Upsilon^2\xi^4}{c^5}\,\omega^5
   + \frac{12\Upsilon^3\xi^6}{c^7}\,\omega^7\,,
\end{equation}
resulting in the 3D phononic VDoS ${\cal D}_{\rm ph}(\omega)$ in Eq.~\eqref{eq:phonon_VDoS} in the main text,
with coefficients
\begin{eqnarray}
    A_{_{\rm D}} & = &\frac{V}{6\pi^2 N}\left(\frac{2}{c_{\rm s}^3} + \frac{1}{c_\ell^3} \right)\,, \nonumber \\
    A_{\rm w} & = & \frac{5\Upsilon V}{6\pi^2 N}\left(\frac{2\xi_{\rm s}^2}{c_{\rm s}^5}+\frac{\xi_{_{\ell}}^2}{c_\ell^5} \right)\,, \\
    A_6 & = & \frac{28\Upsilon^2 V}{6\pi^2N}\left(\frac{2\xi_{\rm s}^4}{c_{\rm s}^7}+\frac{\xi_{_\ell}^4}{c_\ell^7} \right)\,. \nonumber
\end{eqnarray}
We note that including higher-order (${\cal O}(k^5)$ and higher) terms in $\omega(k)$ in Eq.~\eqref{eq:cubic_dispersion_SI} will only affect the ${\cal O}(\omega^6)$ and higher order terms in the phononic VDoS, i.e.~$A_{\rm w}$ is independent of higher order corrections to the dispersion. We generally find the contributions of the $k^5$ terms in the shear and longitudinal dispersion to $A_6$ to be negligible, and therefore do not consider them in our calculations.


\begin{figure}[ht!]
  \includegraphics[width = 0.49\textwidth]{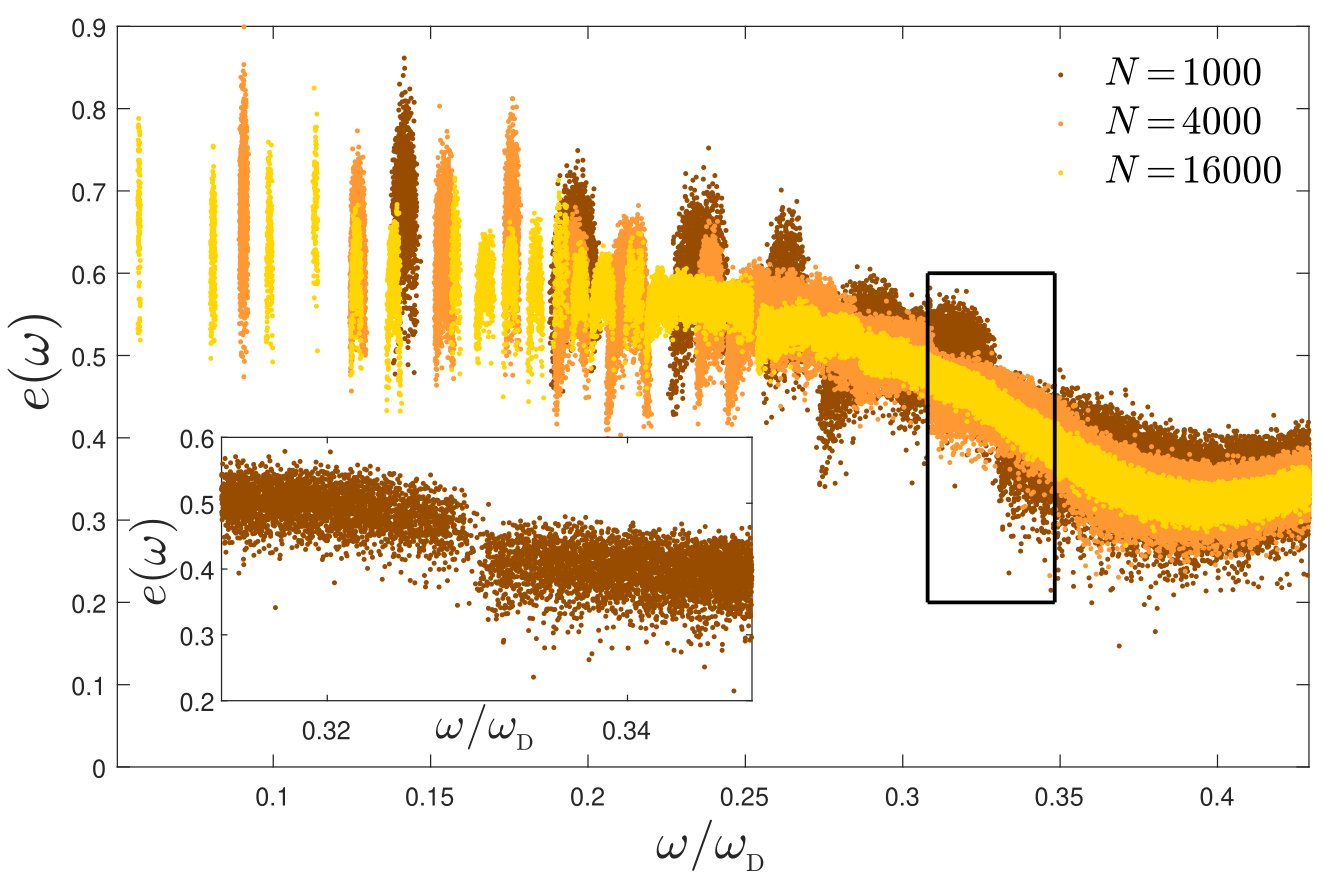}
  \caption{\footnotesize \textbf{Determining the onset frequency $\omega_\star$ of non-phononic vibrations in disordered networks.} The participation ratio $e(\omega)$ of vibrational modes is scatter-plotted against those modes' frequency $\omega$. Data are presented for networks of 3 sizes $N$ as indicated in the legend, with $z\!=\!16.24$ (as in Fig.~3b of the main text). The inset presents a zoom in on the data corresponding to $N\!=\!1000$, where the observed small gap at $\omega/\omega_{_{\rm D}}\!\approx\!0.33$ sets a lower bound on $\omega_\star$, see extensive discussion in the text.}
  \label{fig:omega_star}
\end{figure}

\section{Determination of $\omega_\star$}
\vspace{-0.2cm}
\label{sec:appendix_omega_star}

It is well known that non-phononic vibrations in relaxed disordered spring networks --- such as those employed in this work --- are gapped, i.e., they feature an onset (lower cutoff) frequency $\omega_\star\!\sim\!z\!-\!z_{\rm c}$~\cite{mw_EM_epl}. In Fig.~3 of the main text, we mark the onset frequency $\omega_\star$ of non-phononic vibrational modes by a vertical dash-dotted line. Therein, we have $z\=16.24$ and $\omega_{_{\rm D}}\!\approx\!3.96$ (in simulational units). Here, we explain how $\omega_\star$ is estimated.

To estimate $\omega_\star$, we plot in Fig.~\ref{fig:omega_star} the participation ratio $e$ of vibrational modes vs.~their frequency $\omega$ for disordered networks of 3 sizes $N$, as indicated in the legend. Two important properties of both phononic and non-phononic vibrational modes in such finite-size disordered networks allow us to obtain a lower bound on $\omega_\star$: (i) at low frequencies, phonons are grouped into discrete bands, where the gaps between them decrease with $\omega$ but persist to higher frequencies the smaller $N$ is~\cite{phonon_widths}, and their $e$ decreases with $\omega$ in the range considered. (ii) Non-phononic modes continuously cover the frequency axis (with no gaps) for $\omega\!\gtrsim\!\omega_\star$ and their $e$ is $N$-independent~\cite{ikeda_pnas}. Taken together, one concludes that the largest frequency at which a gap exists in the spectrum for the smallest $N$ is a lower bound on $\omega_\star$.

In the inset of Fig.~\ref{fig:omega_star}, we zoom in on $e(\omega)$ of the smallest $N$ dataset presented in the main panel, where the largest frequency gap is observed. The small (presumably the smallest) gap, appearing at $\omega\!\approx\!0.33\,\omega_{_{\rm D}}$, is also accompanied by a drop in $e(\omega)$, indicating that the modes from both sides of the gap are indeed all phononic, where non-phononic ones emerge only at slightly larger frequencies. Furthermore, we observe that only above $\omega\!\approx\!0.33\,\omega_{_{\rm D}}$ the average participation ratio $e(\omega)$ becomes approximately $N$-independent. We therefore set $\omega_\star\!=\!0.33\,\omega_{_{\rm D}}$, as marked in Fig.~3 of the main text.

\section{Relations to recent work}
\label{sec:NatPhy2025}
\vspace{-0.2cm}

Our findings point to some serious difficulties in recent work. The leading-order nonlinear phonon dispersion relation $\omega(k)\!\simeq\!ck - c\Upsilon\xi^2k^3$ (with $\Upsilon\=1/96$) is presented in the manuscript and in Eq.~\eqref{eq:cubic_dispersion_SI}, featuring a disorder-induced length $\xi$ that satisfies
\begin{equation}
    \xi \gg a_{_0} \ ,
\label{eq:mesoscopic_xi_SI}
\end{equation}
where $a_{_0}$ is an average interparticle distance. We also found a Rayleigh scattering scaling of the wave attenuation rate in the form
\begin{equation}
    \Gamma(k)\!\sim\!\xi^2 k^4 \ ,
\label{eq:Rayleigh_SI}
\end{equation}
for $k\xi\!\ll\!1$.

Recent work~\cite{nature_physics_boson_peak_2025}, see Eq.~(11) therein and note that $q_{_{\rm D}}\=2\pi/a_{_0}$ in the notation therein, claimed that the phonon dispersion relation of glasses takes the form
\begin{equation}
    \omega(k) = \frac{4\,c}{a_{_0}}\sin\!\left(\frac{k\,a_{_0}}{4}\right)\exp\!\left(-\frac{a_{_0}\Gamma(k)}{4\pi c} \right) \ ,
\label{eq:NatPhy2025}
\end{equation}
where the first term identifies with the crystalline (no disorder) phonon dispersion of Eq.~\eqref{eq:dispersion_1D_lattice} and the second (exponential) term embodies the effect of disorder through $\Gamma(k)$. In view of Eq.~\eqref{eq:Rayleigh_SI}, the leading-order nonlinear phonon dispersion relation corresponding to Eq.~\eqref{eq:NatPhy2025} takes the form $\omega(k)\!\simeq\!ck - c\Upsilon\xi^2k^3$ (with $\Upsilon\=1/96$) with $\xi\=a_{_0}$, which contradicts Eq.~\eqref{eq:mesoscopic_xi_SI}.


%

\end{document}